\documentclass[12pt]{article}
\usepackage[utf8]{inputenc}
\usepackage{lmodern}
\usepackage{lineno}
\usepackage{amssymb}
\usepackage{graphicx}
\usepackage{amsfonts}
\usepackage{amsmath}
\usepackage{amsthm}
\usepackage{mathtools}
\usepackage{bm}
\usepackage{color}
\usepackage{pgfplots}
\pgfplotsset{compat=1.16}
\usepackage{here}
\usepackage{multirow}
\usepackage[hang,small,bf]{caption}
\usepackage[subrefformat=parens]{subcaption}
\usepackage{authblk}
\usepackage[colorlinks=true, citecolor=teal, linkcolor=teal, urlcolor=teal]{hyperref}
\usepackage[authoryear]{natbib}
\usepackage{threeparttable}
\usepackage{comment}
\usepackage{stackengine}
\usepackage{tikz}
\usetikzlibrary{arrows.meta}
\usetikzlibrary{bending}
\usepackage{tcolorbox}

\def\E#1{\mathbb{E}\left[#1\right]}
\def\V#1{\mathbb{V}\left[#1\right]}
\def\P#1{\mathbb{P}\left[#1\right]}

\newcommand{\indep}{\perp\!\!\!\!\perp}

\newcommand{\rnabla}{\rotatebox[origin=c]{180}{$\nabla$}}

\numberwithin{equation}{section}

\newtheorem{theoremx}{Theorem}
\newenvironment{theorem}
  {\pushQED{\qed}\theoremx}
  {\popQED\endtheoremx}
\newtheorem{assumption}{Assumption}[section]

\newtheorem{corx}{Corollary}
\newenvironment{corollary}
  {\pushQED{\qed}\corx}
  {\popQED\endcorx}

\newtheorem{lemmax}{Lemma}
\newenvironment{lemma}
  {\pushQED{\qed}\lemmax}
  {\popQED\endlemmax}

\theoremstyle{definition}
\newtheorem{remarkx}{Remark}
 \newenvironment{remark}
  {\pushQED{\qed}\remarkx}
  {\popQED\endremarkx}
\newtheorem{exx}{Example}
\newenvironment{example}
  {\pushQED{\qed}\exx}
  {\popQED\endexx}



\makeatletter
\renewcommand*{\@fnsymbol}[1]{\ensuremath{\ifcase#1\or \flat\or * \else\@ctrerr\fi}}
\makeatother

\title{\bf Robustify and Tighten the Lee Bounds: \\ A Sample Selection Model under Stochastic Monotonicity and Symmetry Assumptions}
\author[$\sharp$]{Yuta Okamoto\footnote{
\href{mailto:okamoto.yuuta.57w@st.kyoto-u.ac.jp}{okamoto.yuuta.57w@st.kyoto-u.ac.jp}.
I gratefully acknowledge financial support from JST SPRING, Grant Number JPMJSP2110.
}}
\affil[$\sharp$]{Graduate School of Economics, Kyoto University}
\date{March 20, 2024}

\begin{document}


\maketitle
\begin{abstract}
    In the presence of sample selection, Lee's (2009) nonparametric bounds are a popular tool for estimating a treatment effect.
    However, the Lee bounds rely on the monotonicity assumption, whose empirical validity is sometimes unclear. Furthermore, the bounds are often regarded to be wide and less informative even under monotonicity.
    To address these issues, this study introduces a stochastic version of the monotonicity assumption alongside a nonparametric distributional shape constraint.
    The former enhances the robustness of the Lee bounds with respect to monotonicity, while the latter helps tighten these bounds.
    The obtained bounds do not rely on the exclusion restriction and can be root-$n$ consistently estimable, making them practically viable.
    The potential usefulness of the proposed methods is illustrated by their application on experimental data from the after-school instruction programme studied by Muralidharan, Singh, and Ganimian (2019).
\end{abstract}


\begin{tcolorbox}[
    colback=blue!5,         
    colframe=blue!80!black, 
    sharp corners,          
    boxrule=1mm,            
    left=2mm,               
    right=2mm,              
    top=2mm,                
    bottom=2mm,             
    fonttitle=\bfseries,    
    fontupper=\large        
]
A revised version of this article has been accepted for publication in \textit{The Econometrics Journal}, published by Oxford University Press. DOI: \href{https://doi.org/10.1093/ectj/utaf001}{10.1093/ectj/utaf001}
\end{tcolorbox}

\newpage

\section{Introduction}\label{sec: introduction}
Sample selection is an important challenge for identifying causal effects in empirical economics (see, e.g., \citealp[p.~881]{Hansen2022econometrics}).
Earlier approaches model the selection mechanism as follows:
\begin{align*}
    Y_i^* = X_i^\prime \beta + \varepsilon_i,\,\,
    Y_i = Y_i^* \mathbf{1}\{W_i^\prime \gamma + \nu_i \geq 0\},
\end{align*}
where $Y_i^*$ is the latent outcome and researchers observe $(Y_i, X_i, W_i)$; and assume that the covariate $W_i$ includes some variables which are not in $X_i$, meaning that researchers can access some ``instrument" that only affects an individual's selection decision but not the outcome.
A similar assumption is made by not only such single index semiparametric models but more general nonparametric models (e.g., \citealp{Das_elat:2003}). 
However, this exclusion restriction is not always satisfied in practice because such an instrument may not be easily accessible. 
To overcome this challenge, \cite{Lee:2009} developed a partial identification strategy that does not necessitate this restriction.
The \citeauthor{Lee:2009} bounds are especially used in field experimental studies, wherein the attrition problem often occurs (e.g., \citealp{Baranov2020, Bursztyn2020, Muralidharan:2019}).
Another application includes \cite{Chen_Roth:2023}.

Although popular, the \citeauthor{Lee:2009} bounds have two important limitations.
The first is the monotonicity assumption.
The identification of the \citeauthor{Lee:2009} bounds relies on the assumption of a monotone response to the treatment assignment; specifically, \textit{for every person}, their outcome under treatment has to be observable if it would be observed in the absence of treatment.
This assumption, while classic and seemingly natural in many contexts, is somewhat strong and its empirical validity can be unclear in several cases, because few economic theories strongly affirm the non-existence of any individuals deviating from behavioural models.

The second limitation is that the \citeauthor{Lee:2009} bounds can be wide. For example, \citet[p.~383]{Mobarak:2023} wrote, ``[c]ommon nonparametric bounds (e.g., \citeauthor{Lee:2009} \citeyear{Lee:2009}) are often wide and uninformative, as they trim the data from the extremes of the outcome distribution." 
Several studies provide a similar assessment (e.g., \citealp{Barrow_Rouse:2018, Delius_Sterck:2024}).
Therefore, considering the \citeauthor{Lee:2009} bounds are the tightest possible under his assumptions, it will be beneficial to consider an additional, reasonable restriction to help draw more informative policy implications.

This article aims to address these issues by introducing new choices of assumptions. We first relax the monotonicity assumption to its stochastic counterpart, which allows a certain stochastic deviation from the monotone response. This stochastic monotonicity assumption operates on the intuition that ``the monotonicity approximately, but perhaps not exactly, holds." Such scenarios will frequently arise in empirical analysis.
Moreover, embracing a stochastic variation to choice aligns more closely with observed experimental evidence (e.g., \citealp{Tversky1969, Agranov_Ortoleva:2017}).
Afterwards, we introduce a nonparametric distributional assumption that helps tighten the bounds. 
Specifically, we propose using a symmetry assumption---which need not be satisfied exactly (Assumption \ref{assumption mean median})---on the density function of $Y_i^*$ under treatment for always-takers, with its motivation explained graphically.
This symmetry assumption restricts the admissible class of trimmed densities, enabling us to avoid \citeauthor{Lee:2009}'s (\citeyear{Lee:2009}) ``extreme" trimming of the outcome distribution, thereby tightening the bounds over the standard \citeauthor{Lee:2009} bounds.

Based on these assumptions, we develop partial identifications and propose simple estimators. Importantly, the bounds are root-$n$ consistently estimable, making the implementation practically feasible. 
The potential usefulness of the bounds is demonstrated using an empirical example.

This article proceeds as follows.
The rest of this section reviews the related literature.
In Section \ref{sec: bounds under SM}, we discuss the identification of the sample selection model under the stochastic monotonicity assumption.
Section \ref{sec: symmetry assumption} introduces the symmetry assumption and establishes identification. Some numerical examples are also provided in these sections.
Estimation and inference procedures are considered in Section \ref{sec: inference}. 
The empirical illustration in Section \ref{sec: empirical} highlights the usefulness of the obtained bounds. Section \ref{sec: conclusion} concludes this article.

\subsection*{Related Literature}
The sample selection problem has been an important issue in economics (\citealp{Heckman:1979}).
\cite{Lee:2009} is a seminal work that derives nonparametric bounds under the monotonicity assumption based on the insight of \cite{Horowitz_Manski:1995}. Similar studies include \cite{Imai:2008} and \cite{Chen_Flores:2015}. 
Based on a similar idea, \cite{Bartalotti_etal:2023} recently considered the identification of the marginal treatment effect in the presence of sample selection, which essentially extends \citeauthor{Lee:2009}'s model.

This study is not the first to focus on the monotonicity assumption.
Relatedly, \cite{Semenova:2023} explored a different kind of generalization of the \citeauthor{Lee:2009} bounds, specifically identifying the bounds assuming that exact monotonicity holds conditional on the covariates.
This approach will be very useful when the direction of selection effects systematically differs depending on the covariates. 
However, in some cases, researchers may not have access to such informative covariates. Besides, even conditional on the covariates, the requirement that the monotone response in selection has to hold for every such person may still be practically strong.
Hence, when the predictive power of the covariates seems weak but one can assume that the fraction of those deviating from complete monotonicity is not so large, our approach can be a useful complementary tool.

Our relaxation of the exact monotonicity assumption using the information of the lower bound on a parameter (see Assumption \ref{assumption: stochastic monotonicity}) is similar to the techniques used in the sensitivity analysis literature, such as \cite{Conley_etal:2012}, who treated the exogeneity condition in linear instrumental variable methods.
The stochastic monotonicity assumption is also similar to the bounded variation assumption found in the difference-in-differences and regression discontinuity literature (e.g., \citealp{Imbens_Wager:2019, Manski_Pepper:2018, Rambachan_Roth:2023}).
Note that, in our setting, the lower bound (denoted by $\vartheta_L$ in Assumption \ref{assumption: stochastic monotonicity}) can be easier for a researcher to set because $\vartheta_L$ is the bound on probability, and thus, has an intuitive meaning. Furthermore, what we are imposing is clear.

The width of the \citeauthor{Lee:2009} bounds are also discussed in the literature.
\cite{Beheghel_etal:2015} proposed an interesting idea to derive narrower nonparametric bounds on the treatment effect, whereas an instrument that only affects the participation decision is needed.
\cite{Honore_Hu:2020, Honore_Hu:2022} derived much sharper bounds than \citeauthor{Lee:2009} bounds under a fully structured semiparametric models.
Although their results do not rely on the excluded instrument, their identification depends on the imposed structural assumptions, meaning that it is less robust to specification error than nonparametric models as \cite{Lee:2009}.
Intuitively, our approach utilising (nonparametric) distributional assumption is inbetween these two studies: Our procedure does not require the additional instruments but still builds on \citeauthor{Lee:2009}'s insight.

Finally, our procedure can be applicable to various empirical settings in which sample selection can occur.
A prominent example is the attrition problem in randomised controlled trials (\citealp{Duflo_etal:2007}, Section 6.4). 
To demonstrate such an application, we use the data from \citeauthor{Muralidharan:2019}'s (\citeyear{Muralidharan:2019}) study as an empirical example in Section \ref{sec: empirical}.

\section{Bounds under Stochastic Monotonicity}\label{sec: bounds under SM}
Here, we introduce a stochastic version of the monotonicity assumption and provide the partial identification under this assumption.
Section \ref{subsec: model} reviews the general selection model of \cite{Lee:2009} and the monotonicity assumption.
In Section \ref{subsec: st monotonicity and identification}, we extend \citeauthor{Lee:2009}'s (\citeyear{Lee:2009}) results to establish the more robust bounds under the stochastic monotonicity.

\subsection{General Selection Model and Monotonicity Assumption}\label{subsec: model}
Following \cite{Lee:2009}, we consider the following general selection model:
\begin{subequations}\label{general model}
\begin{align}
    &(Y_{1i}^*, Y_{0i}^*, S_{1i}, S_{0i}, D_i)\text{ is i.i.d. across individuals},\\
    & S_i=S_{1i}D_i + S_{0i}(1-D_i),\\
    &Y_i = S_i\cdot\left\{Y_{1i}^* D_i + Y_{0i}^* (1-D_i)\right\},\\
    &(Y_i, S_i, D_i)\text{ is observed},
\end{align}
\end{subequations}
where $Y_{1i}^*$ ($Y_{0i}^*$) is the potential outcome when an individual $i$ is assigned to a treatment (control) group; $D_i\in\{0,1\}$ is a binary indicator of treatment status that equals $1$ when the individual is treated; and $S_{1i}\in\{0,1\}$ and $S_{0i}\in\{0,1\}$ are the potential sample selection indicator under the treated and control states, respectively. $S_i$ equalis $1$ when the individual $i$ does not exhibit attrition, and $0$ otherwise.

We are interested in the average treatment effect for the always takers ($S_{0i}=S_{1i}=1$), or 
\begin{align*}
    \tau\coloneqq\E{Y_{1i}^* - Y_{0i}^*|S_{0i}=S_{1i}=1}.
\end{align*}
\cite{Lee:2009} showed that $\tau$ can be partially identified under the following assumptions:
\begin{assumption}[Independent Assignment]\label{assumption: independence}
    $(Y_{1i}^*, Y_{0i}^*, S_{1i}, S_{0i})$ is independent of $D_i$.
\end{assumption}
\begin{assumption}[Monotonicity]\label{assumption: monotonicity}
    $\P{S_{1i}=1 | S_{0i}=1}=1$.
\end{assumption}

The basic idea of the \citeauthor{Lee:2009} bounds is as follows.
Under Assumptions \ref{assumption: independence} and \ref{assumption: monotonicity}, we have
\begin{align*}
    \tau = \E{Y_{1i}^*|S_{0i}=S_{1i}=1} - \E{Y_{i}|D_i=0,S_i=1}.
\end{align*}
The point identification fails due to the first term of the right-hand side, as it is not identified. 
Here, under Assumption \ref{assumption: monotonicity}, we have the following decompsition:
\begin{align}
    &\underbrace{\P{Y_i\leq y|D_i=1, S_i=1}}_{\text{identified}}\notag\\
    &\qquad=
     \P{Y_{1i}^*\leq y|S_{0i}=0, S_{1i}=1} (1-q_0)
     +\underbrace{\P{Y_{1i}^*\leq y|S_{0i}=1, S_{1i}=1}}_{\text{our interest}} q_0,\label{lees decomposition}
\end{align}
where $q_0$ is defined in Theorem \ref{thm identification}, which is identified and in $[0,1]$ under Assumption \ref{assumption: monotonicity}.
By assuming that the smallest $q_0$ values of $Y_i$ are entirely attributed to the group with $S_{0i}=1, S_{1i}=1$, we can derive the lower bound as follows:
\begin{align*}
    \E{Y_{1i}^*|S_{0i}=S_{1i}=1} \geq \underbrace{\E{Y_i|D_i=1, S_i = 1, Y_i\leq y^1_{q_0}}}_{\text{identified}},
\end{align*}
which is \citeauthor{Lee:2009}'s lower bound. 
The upper bound can be obtained similarly.

Although it plays several important roles in the above transformation, the monotonicity assumption is somewhat restrictive. 
Intuitively speaking, it requires that \textit{every individual} in the population obeys the behavioural rule that $S_{0i}=1\Longrightarrow S_{1i}=1$ \textit{with no exception}. This may fail if the population consists of heterogeneous individuals.
Further, as discussed in Section \ref{sec: empirical} using an empirical example, several context-dependent concerns about its plausibility may exist.

However, if the monotonicity assumption is roughly true, would the \citeauthor{Lee:2009} bounds successfully produce valid bounds?
Unfortunately, a slight deviation from the exact monotonicity can lead to misleading conclusions, as illustrated in the next example:

\begin{example}\label{ex 1}
$D_i$ is assigned randomly.
On the selection mechanism, suppose 
\begin{align*}
    \begin{cases}
        \P{S_{1i} = 1, S_{0i} = 1} = 0.475,
        &\P{S_{1i} = 0, S_{0i} = 1} = 0.025,\\
        \P{S_{1i} = 1, S_{0i} = 0} = 0.300,
        &\P{S_{1i} = 0, S_{0i} = 0} = 0.200.
    \end{cases}
\end{align*}
Thus, a small fraction (2.5\%) of the population deviates from the monotone response.
On the outcome, suppose that the conditional densities are given by
\begin{align*}
    \begin{cases}
        f_{Y_{0}^*}(y | S_{1i}=k, S_{0i}=1)=\phi(y;0,1/2),
        &f_{Y_{1}^*}(y | S_{1i}=k, S_{0i}=1)=\phi(y;0,1/2),\\
        f_{Y_{0}^*}(y | S_{1i}=1, S_{0i}=0)=\phi(y; 2,1/2),
        &f_{Y_{1}^*}(y | S_{1i}=1, S_{0i}=0)=\phi(y; 2,1/2),
    \end{cases}
\end{align*}
where $k\in\{0,1\}$ and $\phi(\cdot;\mu,\sigma^2)$ is the density function of the normal distribution $\mathcal{N}(\mu, \sigma^2)$. 
The treatment has no effects for always takers, i.e., $\tau=0$.
The \citeauthor{Lee:2009} bounds under the monotonicity assumption are computed as $[0.02, 1.46]$, which do not cover the true $\tau$. 
\end{example}

Therefore, when the monotonicity does not exactly hold, the \citeauthor{Lee:2009} bounds can be invalid.

\subsection{Stochastic Monotonicity and Partial Identification}\label{subsec: st monotonicity and identification}
When computing the \citeauthor{Lee:2009} bounds, economists often operate under the premise that the monotonicity assumption is largely, though perhaps not perfectly, met.
As highlighted by the above example, considering bounds that account for this nuanced understanding can enhance robustness.
To align with this perspective, this subsection introduces a weaker and interpretable assumption that builds on the intuition that ``the monotonicity mostly, but perhaps not exactly, holds" and then derives the bounds on the treatment effect $\tau$ under this assumption.

In particular, we consider the identification under the following stochastic version of the monotonicity:
\begin{assumption}[Stochastic Monotonicity]\label{assumption: stochastic monotonicity}
    \item[(i)] $\P{S_{1i}=1|S_{0i}=1}\geq\vartheta_L$ with known $\vartheta_L\in(0,1]$.
    \item[(ii)] $(Y_{1i}^*, Y_{0i}^*)$ is independent of $S_{1i}$ conditional on $S_{0i}=1$.
\end{assumption}
\begin{remark}
    Assumption \ref{assumption: monotonicity} is a special case of Assumption \ref{assumption: stochastic monotonicity} with $\vartheta_L=1$.
\end{remark}

Part (i) states that the monotonicity holds with a probability of no less than some known value $\vartheta_L$. 
This $\vartheta_L$ can be considered a representation of the validity of the monotonicity.
It can be determined based on a researcher's institutional knowledge and beliefs. For example, when a researcher assumes that almost all participants satisfy the monotonicity assumption but anticipates that a small fraction of them perhaps deviates from the monotonic behaviour, setting $\vartheta_L=0.95$ (alternatively $\vartheta_L=0.90$ or $0.99$) is reasonable in a similar vein to the statistical significance level. 
This kind of bound on a parameter is common in various settings (\citealp{Conley_etal:2012, Manski_Pepper:2018, Imbens_Wager:2019, Rambachan_Roth:2023}), whereas its intuitive meaning is not always clear.
However, the lower bound $\vartheta_L$ has a clear meaning and it is easy to map the researcher's knowledge to it.

The lower bound $\vartheta_L$ provides another intuitive relation. 
Under Assumption \ref{assumption: independence}, we have
\begin{align}
    \P{S_{1i}=1 | S_{0i}=0} = \frac{\P{S_i=1 | D_i=1} - \P{S_{1i}=1|S_{0i}=1}\P{S_i=1 | D_i=0}}{1 - \P{S_i=1 | D_i=0}}.\label{vartheta lower bound}
\end{align}
Hence, under Assumption \ref{assumption: stochastic monotonicity}-(i), it follows that
\begin{align}
    \P{S_{1i}=1 | S_{0i}=0} \leq \frac{\P{S_i=1 | D_i=1} - \vartheta_L\P{S_i=1 | D_i=0}}{1 - \P{S_i=1 | D_i=0}}.\label{P upper}
\end{align}
That is, setting $\vartheta_L$ is equivalent to determining the upper bound on $\P{S_{1i}=1 | S_{0i}=0}$ once the identified parameters are known. This could be used to empirically validate Assumption \ref{assumption: stochastic monotonicity}-(i) by computing the empirical analogue of the upper bound and checking if $\P{S_{1i}=1 | S_{0i}=0}$ is not unreasonably small.

Part (ii) states that the attrition and potential outcomes are irrelevant for those whose outcomes are (potentially) observable without treatment.
\citeauthor{Lee:2009}'s monotonicity assumption automatically satisfies this. However, it is not explicitly treated in \cite{Lee:2009}, and thus, we discuss this assumption a little more.
Assumption \ref{assumption: stochastic monotonicity}-(ii) can be rewritten as
\begin{align}
    \P{S_{1i} = 1 | Y_{1i}^*, Y_{0i}^*, S_{0i}=1} = \P{S_{1i} = 1 | S_{0i}=1}.\label{sto mon eq}
\end{align}
Importantly, $Y_{1i}^*, Y_{0i}^*$ can affect $S_{1i}$ only through $S_{0i}$ if $S_{0i}=1$. When we suppose that $S_{1i}$ still depends on $Y_{1i}^*, Y_{0i}^*$ after controlling $S_{0i}=1$, condition \eqref{sto mon eq} will fail.
In contrast, this condition can be satisfied, for example, when
\begin{align}
    S_{0i} = \mathbf{1}\{k_0(Y_{1i}^*, Y_{0i}^*) \geq \zeta_{0i}\},\,
    S_{1i} = \mathbf{1}\{k_1(S_{0i}) \geq \zeta_{1i}\}\text{ for those }S_{0i}=1,\label{ok case}
\end{align}
where $k_0$ and $k_1$ are some structural functions, and $\zeta_{0i}$ and $\zeta_{1i}$ are idiosyncratic errors. 
This behavioural structure includes the following examples:
\begin{example}\label{ex1 of job training}
    Consider an off-the-job training programme for employees and its effect on earnings. $D_i$ is the treatment indicator of the job-training, and $Y_{1i}^*, Y_{0i}^*$ are earnings. $S_{0i}, S_{1i}$ denote labour force participation.
    Assume that the decision-makers participate in the labour market only if their subjectively expected earnings are above the predetermined reservation wage $r_i(>0)$; $Y_{di}^*$ is decomposed as the sum of the deterministic baseline ($b_i$) and stochastic performance-based pay ($p_{di}$), i.e., $Y_{di}^* = b_i + p_{di}$; and $\mathbb{E}^{\mathrm{sbj}}_{i}\left[p_{1i}\right] = \mathbb{E}^{\mathrm{sbj}}_{i}\left[p_{0i}\right] + e_i$, where $e_i\geq0$ and $\mathbb{E}^{\mathrm{sbj}}_{i}$ denotes the subjective expectation.
    This may be likely when the potential benefit is explained by the programme staff.
    In this setting, for those $S_{0i}=1$, it follows that
    \begin{align*}
        S_{0i} &= \mathbf{1}\{\mathbb{E}^{\mathrm{sbj}}_{i}\left[Y_{0i}^*\right] \geq r_i\} = 
        \mathbf{1}\{b_i + \mathbb{E}^{\mathrm{sbj}}_{i}\left[p_{0i}\right] \geq r_i\},\\
        S_{1i} &= \mathbf{1}\{\mathbb{E}^{\mathrm{sbj}}_{i}\left[Y_{1i}^*\right] \geq r_i\} = 
        \mathbf{1}\{b_i + \mathbb{E}^{\mathrm{sbj}}_{i}\left[p_{0i}\right] + e_i \geq r_i\} = \mathbf{1}\{S_{0i} =1\}\text{ for those }S_{0i}=1,
    \end{align*}
    This is a special case of \eqref{ok case}. Hence, Assumption \ref{assumption: stochastic monotonicity}-(ii) is satisfied. Assumption \ref{assumption: monotonicity} also holds.
\end{example}

\begin{example}\label{ex3 of job training}
    Consider again the situation in Example \ref{ex1 of job training}.
    Assume here that the job training requires some utility cost $c_i\in\{0, \infty\}$ that is independent of $(Y_{1i}^*, Y_{0i}^*)$, $0<\P{c_i=\infty}\ll 1$.
    The setup is a simplification made to indicate that, for the most part, the utility cost is sufficiently small.
    Now, suppose that
    \begin{align*}
        S_{0i} = \mathbf{1}\{\mathbb{E}^{\mathrm{sbj}}_{i}\left[Y_{0i}^*\right] \geq r_i\},\,
        S_{1i} = \mathbf{1}\{\mathbb{E}^{\mathrm{sbj}}_{i}\left[Y_{1i}^*\right] - c_i\geq r_i\}.
    \end{align*}
    This may happen when the job training programme is held at night after clocking out and can be physically/mentally very hard for someone; $c_i=\infty$ indicates that the programme is so hard for individual $i$ that, say, goes on sick leave.
    Then,
    \begin{align*}
        S_{0i} &= 
        \mathbf{1}\{b_i + \mathbb{E}^{\mathrm{sbj}}_{i}\left[p_{0i}\right] \geq r_i\},\\
        S_{1i} &= 
        \mathbf{1}\{b_i + \mathbb{E}^{\mathrm{sbj}}_{i}\left[p_{0i}\right] + e_i - c_i \geq r_i\} = \mathbf{1}\{S_{0i}\times\mathbf{1}\{c_i=0\} =1\}\text{ for those }S_{0i}=1,
    \end{align*}
    which is a special case of \eqref{ok case}.
    In this case, Assumption \ref{assumption: stochastic monotonicity}-(ii) holds, but Assumption \ref{assumption: monotonicity} fails.
    If we can assume $\P{c_i<\infty}\geq \vartheta_L$ with some $\vartheta_L$, Assumption \ref{assumption: stochastic monotonicity}-(i) holds.
\end{example}

As illustrated in Example \ref{ex3 of job training}, Assumption \ref{assumption: stochastic monotonicity} accounts for scenarios where Assumption \ref{assumption: monotonicity} is largely reasonable, yet subject to potential deviations due to \textit{stochastic shocks} affecting the decision-makers' tastes.
Note that, other than this rather restrictive model's setup in Example \ref{ex3 of job training}, such a stochastic flip of choice is widely acknowledged in decision theory, as evidenced by \cite{Tversky1969} and \cite{Agranov_Ortoleva:2017}.

Under this stochastic monotonicity assumption, we have the following bounds.
\begin{theorem}\label{thm identification}
    Suppose the model \eqref{general model}.
    Let $Y_{0i}^*$ and $Y_{1i}^*$ be continuous random variables. Under Assumptions \ref{assumption: independence} and \ref{assumption: stochastic monotonicity}, the sharp bounds $[\rnabla, \nabla]$ are given by
    \begin{align*}
        \nabla &= \E{Y_i|D_i=1, S_i = 1, Y_i\geq y^1_{1 - \vartheta q_0}} - \E{Y_i|D_i=0,S_i=1},\\
        \rnabla &= \E{Y_i|D_i=1, S_i = 1, Y_i\leq y^1_{\vartheta q_0}} - \E{Y_i|D_i=0,S_i=1},
    \end{align*}
    where $y^1_r = \inf\{t\in\mathbb{R} : r \leq F_1(t)\}$ with $F_1$ the cumulative distribution function of $Y_i$ conditional on $D_i=1, S_i=1$, $q_0 = {\P{S_i=1|D_i=0}}/{\P{S_i=1|D_i=1}}$, $\vartheta=\max\{\vartheta_L,\vartheta_F\}$, $\vartheta_F = 1 + 1/q_0 - 1/\alpha_0$, and $\alpha_0=\P{S_i=1|D_i=0}$.
\end{theorem}

The max operator and $\vartheta_F$ are due to the Fréchet inequality, which is equivalent to the inequality constraint $\P{S_{1i}=1|S_{0i}=0}\leq 1$; see also \eqref{vartheta lower bound}. This ensures that $\P{S_{1i}=1|S_{0i}=1}\geq\vartheta_F$ under Assumption \ref{assumption: independence}. 
Therefore, the trimming is not performed at $y^1_{\vartheta_L q_0}$ but at $y^1_{\vartheta q_0}$.

When $\vartheta_L=1$, the bounds coincide with \citeauthor{Lee:2009}'s (\citeyear{Lee:2009}).
As can be seen from Theorem \ref{thm identification}, the length between the bounds becomes wider as $\vartheta_L$ decreases as long as $\vartheta_L\geq\vartheta_F$.
This is natural since the less information we have about the individual's selection rule, the greater the length of the bounds.
With this monotonic relation between $\vartheta_L$ and the bounds, the bounds under Assumption \ref{assumption: stochastic monotonicity} can also be used to perform sensitivity analysis or examine the monotonicity assumption's identifying power. 
For example, when researchers are concerned about whether the monotone response assumption is reasonable, drawing the bounds with multiple $\vartheta_L$ will be helpful.
If the bounds do not cover zero with somewhat smaller $\vartheta_L$, researchers can safely conclude that the treatment has a positive effect.

In the previous subsection, we saw a simple example where a small deviation from the monotonicity leads to misleading bounds. 
The stochastic monotonicity can produce more robust bounds:
\begin{example}[Example \ref{ex 1} Continued]\label{ex 2}
    Assume the stochastic monotonicity assumption holds with $\vartheta_L=0.95$. 
    Then, $\vartheta=0.95$ as $\vartheta_F = 0.55$.
    In contrast to the bounds obtained in Example \ref{ex 1}, the bounds under stochastic monotonicity are $[-0.04, 1.53]$, which correctly covers zero (true effect).
\end{example}

While the bounds have been made more robust with respect to the deviation from monotonicity, this robustification may lead to wider bounds that could obscure policy implications. 
This is especially concerning given that the standard \citeauthor{Lee:2009} bounds are already considered to be wide. 
The subsequent section explores a method to tighten these bounds while maintaining robustness against deviations from monotonicity.

\section{Tightening Bounds using Symmetry Assumption}\label{sec: symmetry assumption}
\subsection{Motivation and Idea}
The \citeauthor{Lee:2009} bounds are often argued to be too wide and less informative even under the exact monotonicity assumption (e.g., \citealp{Barrow_Rouse:2018, Delius_Sterck:2024, Mobarak:2023}).
To address this practical issue, we introduce an additional assumption to tighten the bounds. As this assumption does not require additional variables that satisfy the exclusion restriction, the following identification result can be useful in the absence of such variables. With such an additional excluded variable at hand, the procedure proposed in \cite{Beheghel_etal:2015} may be an alternative tool.

\begin{figure}[t]
    \centering
    \begin{subfigure}[b]{0.49\textwidth}
        \centering
        \includegraphics[scale=0.6]{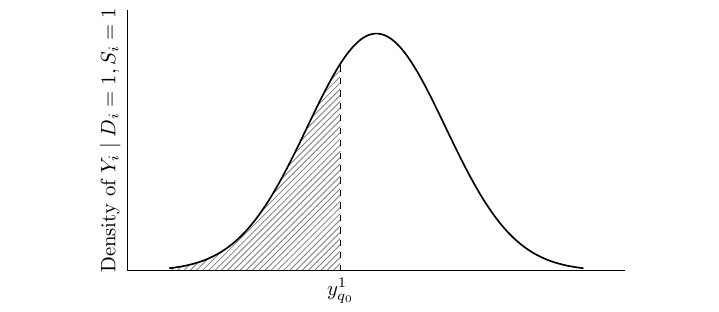}
        \caption{Trimmed Density of Lee's Procedure}
        \label{fig: tikz1}
    \end{subfigure}
    \hfill
    \begin{subfigure}[b]{0.49\textwidth}
        \centering
        \includegraphics[scale=0.6]{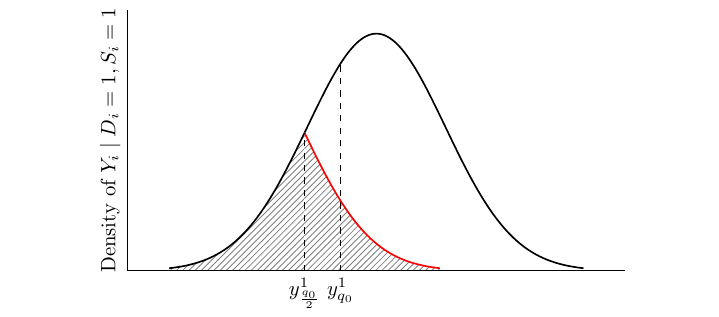}
        \caption{Trimmed Density under Symmetry}
        \label{fig: tikz2}
    \end{subfigure}
    \caption{Comparison of Trimmed Densities. \textit{Notes: The shaded area in (b) is constructed by folding the solid line to the left of $y^1_{q_0/2}$ toward the right.}}
\end{figure}

Before introducing the assumption, we again recall the idea behind the \citeauthor{Lee:2009} bounds. This clarifies why the bounds can be wide and motivates our assumption as a natural extension.
For ease of exposition, temporarily assume that the exact monotonicity holds.
As noted in Section \ref{subsec: model}, the point identification fails due to $\E{Y_{1i}^*|S_{0i}=S_{1i}=1}$.
An important insight from \cite{Lee:2009} was the decomposition \eqref{lees decomposition}, and we obtain \citeauthor{Lee:2009}'s lower bound as
\begin{align*}
    \E{Y_{1i}^*|S_{0i}=S_{1i}=1} \geq \E{Y_i|D_i=1, S_i = 1, Y_i\leq y^1_{q_0}},
\end{align*}
\textit{by assuming that the smallest $q_0$ values of $Y_i$ is entirely attributed to $Y_{1i}^*$ of the group with $S_{0i}= S_{1i}=1$}.
Importantly, this is equivalent to assuming that the conditional density of $Y_{1i}^*$ given $S_{0i}=S_{1i}=1$ can, after appropriate rescaling, have the form like the shaded area in Figure \ref{fig: tikz1}, wherein the solid line represents the density of $Y_i$ conditional on $D_i=1, S_i=1$.

Is this form, right-truncated at $y^1_{q_0}$, empirically natural?
In several cases, this may be counter-intuitive and perhaps makes practitioners suppose that ``\citeauthor{Lee:2009} Bounds are based on extreme assumptions about sample selection" (\citealp{Delius_Sterck:2024}).

In many empirical contexts, a ``well-behaved" density is more probable. 
One promising candidate is the class of symmetric densities, especially since the distribution of several economically important variables, such as log wages and test scores, frequently shows approximate symmetry (see also Section S3 of the Online Appendix and Figure \ref{fig: KDE sharpness}).
Assuming the conditional density is symmetric, we can avoid the aforementioned right-truncated trimmed density, and the lower bound is attained when it has the form as in Figure \ref{fig: tikz2}.
Such a functional form may not be counter-intuitive and perhaps more natural in practice. 
Additionally, this additional information could tighten the bounds.
Further, the symmetry immediately implies that the lower bound is given by
\begin{align*}
    \E{Y_{1i}^*|S_{0i}=S_{1i}=1} \geq y^1_{\frac{q_0}{2}},
\end{align*}
which can be easily estimated, bypassing the need for, for example, a first-stage nonparametric estimation.
The following subsection formalises the symmetry assumption and provides identification. 

\subsection{Assumptions and Partial Identification}\label{subsec: symmetry assumptions}

Motivated by the previous discussion, we introduce the following symmetry assumption:
\begin{assumption}[Symmetry]\label{assumption: symmetry}
    The conditional density function of $Y_{1i}^*$ given $S_{0i}=S_{1i}=1$ is a symmetric function.
\end{assumption}

We also introduce the following tail smoothness condition, which is implicitly used in Figure \ref{fig: tikz2}. This assumption is not essential to obtain
valid bounds.
\begin{assumption}[Tail Smoothness]\label{assumption: tail}
    Let $f_{y}^1$ be the density function of $Y_i$ conditional on $D_i=1$ and $S_i=1$.
    Given $\vartheta$ and $q_0$ (defined in Theorem \ref{thm identification}), for any $a\in\mathbb{R}_{+}$,
    \begin{align*}
        f_{y}^1\left(y_{\frac{\vartheta q_0}{2}}^1 - a\right) \leq f_{y}^1\left(y_{\frac{\vartheta q_0}{2}}^1 + a\right)\text{ and }
        f_{y}^1\left(y_{1-\frac{\vartheta q_0}{2}}^1 - a\right) \geq 
        f_{y}^1\left(y_{1-\frac{\vartheta q_0}{2}}^1 + a\right).
    \end{align*}
\end{assumption}

Intuitively, it says that the shaded area in Figure \ref{fig: tikz2} is below the conditional density $f_{y}^1$.
Before discussing these assumptions, we provide the identification results:
\begin{theorem}\label{thm identification under symmetry}
    Suppose the model \eqref{general model}.
    Let $Y_{0i}^*$ and $Y_{1i}^*$ be continuous random variables. Assume Assumptions \ref{assumption: independence}, \ref{assumption: stochastic monotonicity}, and \ref{assumption: symmetry} holds. Then, $[\rnabla^{\mathrm{s}}, \nabla^{\mathrm{s}}]$ are valid bounds, i.e., $\tau\in[\rnabla^{\mathrm{s}}, \nabla^{\mathrm{s}}]$, where
    \begin{align*}
        \nabla^{\mathrm{s}} &= y^1_{1 - \frac{\vartheta q_0}{2}} - \E{Y_i|D_i=0,S_i=1},\\
        \rnabla^{\mathrm{s}} &= y^1_{\frac{\vartheta q_0}{2}} - \E{Y_i|D_i=0,S_i=1},
    \end{align*}
    and $y^1_r$, $q_0$, and $\vartheta$ are defined in Theorem \ref{thm identification}.
    Besides, if Assumption \ref{assumption: tail} is also true, the lower and upper bounds, $\rnabla^{\mathrm{s}}$ and $\nabla^{\mathrm{s}}$, are attainable; that is, there can be some joint distributions of $(Y_{1i}^*, Y_{0i}^*, S_{1i}, S_{0i}, D_i)$ that are consistent with the assumptions, induce distribution of $(Y_i, S_i, D_i)$, and satisfy $\tau=\rnabla^{\mathrm{s}}$ or $\tau=\nabla^{\mathrm{s}}$.
\end{theorem}
\begin{corollary}\label{cor: narrower}
    Under the assumptions stated in Theorems \ref{thm identification} and \ref{thm identification under symmetry}, $[\rnabla^{\mathrm{s}}, \nabla^{\mathrm{s}}]\subseteq [\rnabla, \nabla]$.
\end{corollary}

Therefore, we can obtain tightened bounds.
Besides, as Theorem \ref{thm identification under symmetry} suggests, the symmetry assumption can be used in conjunction with the stochastic monotonicity assumption, as illustrated by the following example:
\begin{example}[Examples \ref{ex 1} and \ref{ex 2} Continued]
Recall Examples \ref{ex 1} and \ref{ex 2}, in which the monotonicity fails. 
The bounds under stochastic monotonicity with $\vartheta_L=0.95$ and symmetry are $[-0.0026, 1.49]$, which are valid and sharper than the bounds obtained in Example \ref{ex 2}.
\end{example}

We now shift our focus to the assumptions. The symmetry assumption might provoke the most debate. Given the common use of symmetric distributions in economic modelling and econometrics (e.g., \citealp{Powell:1986}), this assumption is likely to be accepted among economists and serves as a reasonable starting point.
Moreover, some economically important variables, such as log wages, seem to often exhibit symmetric-like distributions, as documented in the Online Appendix (Section S3). 
Furthermore, exact adherence to the symmetry assumption is not requisite. Indeed, the validity of the bounds, which is the most important part in practice, can still be ensured under a less restrictive condition; that is, the mean-median coincidence assumption:
\begin{assumption}[Mean-Median Coincidence]\label{assumption mean median}
    \begin{align*}
        \mathrm{Median}\left[{Y_{1i}^*|S_{0i}=S_{1i}=1}\right] = \E{Y_{1i}^*|S_{0i}=S_{1i}=1}.
    \end{align*}
\end{assumption}
Formally, Theorem \ref{thm identification under symmetry} holds by assuming this instead of Assumption \ref{assumption: symmetry}.
Therefore, a much larger class of distributions, including asymmetric distributions, is allowed for $[\rnabla^{\mathrm{s}}, \nabla^{\mathrm{s}}]$ to be valid. 

Further justification is contingent on the specific empirical contexts and the researchers' expertise. An illustrative discussion on this using a concrete empirical example is provided in Section \ref{subsec: validity}.

Note that the set of Assumptions \ref{assumption: symmetry} to \ref{assumption mean median} can be viewed as an additional layer for a layered analysis, which examines how the bounds differ with our imposed assumptions (see \citealp{Manski_Nagin:1998}). 
Therefore, reporting the bounds with and without the symmetry assumption is valuable, even amidst certain scepticism. 
This enables not just researchers but also policymakers to reassess the credibility of this assumption, and adjust their interpretation of the reported bounds accordingly.

Second, an empirically important result is Corollary \ref{cor: narrower}. 
For this result, Assumption \ref{assumption: tail} plays a role.
Meanwhile, this assumption is just a sufficient condition for $[\rnabla^{\mathrm{s}}, \nabla^{\mathrm{s}}]\subseteq [\rnabla, \nabla]$ to be true.
For example, $\rnabla\leq\rnabla^{\mathrm{s}}$ is often satisfied when the trimmed density without the symmetry assumption (i.e., the shaded area of Figure \ref{fig: tikz1}) is left-skewed, as is the case in Figure \ref{fig: tikz1}.
The left-skewness often occurs when the density smoothly decreases in its left tail, due to \citeauthor{Lee:2009}'s trimming procedure (Figure \ref{fig: tikz1}).
Hence, in many cases, $[\rnabla^{\mathrm{s}}, \nabla^{\mathrm{s}}]$ may be tightened even when Assumption \ref{assumption: tail} does not hold.

Finally, we can verify Assumption \ref{assumption: tail} by drawing a density estimate using a standard estimation strategy, such as the kernel density estimator or the local polynomial density estimator (\citealp{Cattaneo_etal:2020}). 
This visual inspection is useful to check the sharpness of the bounds, or examine the potential for further narrowing the obtained bounds.
Some discussion on this is provided in Section \ref{sec: empirical} using an empirical example.

\section{Estimation and Inference}\label{sec: inference}
We consider the estimation and inference procedures.
We focus on the lower bound obtained in Theorem \ref{thm identification under symmetry} to avoid redundancy. 
The upper bound is analogous and the procedures for the bounds obtained in Theorem \ref{thm identification} are similar, which are provided in the Online Appendix.
We will consider two types of estimators in the subsequent subsections.

\subsection{When $\vartheta_L \geq \vartheta_F$ is Known}\label{subsec: known}
We first consider when $\vartheta_L \geq \vartheta_F$ is known.
The inequality $\P{S_{1i}=1|S_{0i}=1} \geq \vartheta_F$, or equivalently $\P{S_{1i}=1|S_{0i}=0}\leq 1$, is not a very strong requirement. Hence, $\vartheta_F$ may not be very large. Thus, $\vartheta_L \geq \vartheta_F$ can be satisfied in many applications, especially when $\vartheta_L$ is large, e.g., $\vartheta_L=0.95$.
\citeauthor{Lee:2009}'s exact monotonicity ($\vartheta_L=1$) is an leading example of this case.

Define $\bm{\beta} \coloneqq (\beta^L, q, \alpha, \eta)$ and $\bm{\beta}_0 \coloneqq (\beta^L_0, q_0, \alpha_0, \eta_0)$, where $\beta^L_0 = y^1_{\vartheta_L q_0/2}$ and $\eta_0 = \E{Y_i | D_i=0, S_i=1}$.
Similarly to \citet[p.~1099]{Lee:2009}, defining
\begin{align*}
    g(\bm{\beta}) = 
    \begin{pmatrix}
        \left(\mathbf{1}\left\{Y_i > \beta^L\right\} - \left(1-\dfrac{\vartheta_L q}{2}\right)\right) S_i D_i\\
        \left(q S_i - \alpha\right) D_i\\
        (S_i - \alpha)(1 - D_i)\\
        (Y_i - \eta) S_i (1 - D_i)
    \end{pmatrix},
\end{align*}
we can estimate $\bm{\beta}_0$ by $\min_{\bm{\beta}} (\sum_{i=1}^n g(\bm{\beta}))^\prime(\sum_{i=1}^n g(\bm{\beta}))$.
Write the minimiser by $\hat{\bm{\beta}} = (\hat{\beta}^L, \hat{q}, \hat{\alpha}, \hat{\eta})$. Then, we can estimate the lower bound by $\widehat{\rnabla^{\mathrm{s}}_1} = \hat{\beta}^L - \hat{\eta}$.
We have the next asymptotic normality.
\begin{corollary}\label{cor normality under symmetry}
    Assume that $Y_i$ have bounded support, $f^1_y$ is continuous and strictly positive, $\P{S_i=1|D_i=0}>0$, $\P{S_{1i}>S_{0i}}>0$, and Assumptions \ref{assumption: independence} and \ref{assumption: stochastic monotonicity} hold.
    Then $\sqrt{n}\left(\widehat{\rnabla^{\mathrm{s}}_1}  - \rnabla^{\mathrm{s}}\right) \to_d \mathcal{N}(0, \Omega_L + \Omega_C)$, where
    \begin{align*}
         \Omega_{L} &= \left\{f^1_y\left(y^1_{\frac{\vartheta q_0}{2}}\right)^2\right\}^{-1} \frac{1}{\P{S_i=1,D_i=1}}
        \left\{
        \frac{\vartheta q_0}{2}\left(1-\frac{\vartheta q_0}{2}\right) + \frac{\vartheta^2 q_0^2}{4} \Psi_1 + \frac{\vartheta^2}{4} \Psi_2
        \right\},\\
        \Psi_1 &= \frac{\alpha_0}{q_0}\left(1-\frac{\alpha_0}{q_0}\right) \dfrac{\P{D_i=1}}{\P{S_i=1, D_i=1}},\,
        \Psi_2 = \alpha_0(1-\alpha_0) \left(\frac{\P{D_i=1}}{\P{D_i=0}}\right)^2 \dfrac{\P{D_i=0}}{\P{S_i=1, D_i=1}},
     \end{align*}
     and $\Omega_C = \V{Y_i| S_i=1, D_i=0}/\P{S_i=1, D_i=0}$.
\end{corollary}

We can define $\widehat{\nabla^{\mathrm{s}}_1} $ similarly and have $\sqrt{n}\left(\widehat{\nabla^{\mathrm{s}}_1}  - \nabla^\mathrm{s}\right) \to_d \mathcal{N}(0, \Omega_U + \Omega_C)$, where 
\begin{align*}
    \Omega_{U} &= \left\{f^1_y\left(y^1_{1 - \frac{\vartheta q_0}{2}}\right)^2\right\}^{-1} \frac{1}{\P{S_i=1,D_i=1}}
        \left\{
        \frac{\vartheta q_0}{2}\left(1-\frac{\vartheta q_0}{2}\right) + \frac{\vartheta^2 q_0^2}{4} \Psi_1 + \frac{\vartheta^2}{4} \Psi_2
        \right\}
\end{align*}
The probabilities and conditional variance can be estimated by their sample analogues. The conditional density $f^1_y$ can be consistently estimated by the standard kernel method, and standard bandwidth selectors can be used by assuming additional smoothness assumption on $f^1_y$ (see, e.g., \citealp{Jones_etal:1996}).
Then, the asymptotic variances can be consistently estimated by the continuous mapping theorem.
Therefore, similarly to \cite{Lee:2009}, we can perform inference based on \citeauthor{Imbens_Manski:2004}'s (\citeyear{Imbens_Manski:2004}) confidence interval (CI).

\subsection{Unknown Case}\label{subsec: unknown}
Next, we consider the estimation and inference procedure when a researcher does not know which is larger, $\vartheta_L$ and $\vartheta_F$.
Define $\bm{\gamma} \coloneqq (\gamma^L, \gamma^F, q, \alpha, \eta)$ and $\bm{\gamma}_0 \coloneqq (\gamma^L_0, \gamma^F_0, q_0, \alpha_0, \eta_0)$, where $\gamma^L_0 = y^1_{\vartheta_L q_0/2}$ and $\gamma^F_0 = y^1_{(1 + q_0(1-1/\alpha_0))/2}$.
Defining
\begin{align*}
    \Tilde{g}(\bm{\gamma}) = 
    \begin{pmatrix}
        \left(\mathbf{1}\left\{Y_i > \gamma^L\right\} - \left(1-\dfrac{\vartheta_L q}{2}\right)\right) S_i D_i\\
        \left(\mathbf{1}\left\{Y_i > \gamma^F\right\} - \left(1-\dfrac{(1 + q(1-1/\alpha))}{2}\right)\right) S_i D_i\\
        \left(q S_i - \alpha\right) D_i\\
        (S_i - \alpha)(1 - D_i)\\
        (Y_i - \eta) S_i (1 - D_i)
    \end{pmatrix},
\end{align*}
we can estimate $\bm{\gamma}_0$ by $\min_{\bm{\gamma}} (\sum_{i=1}^n \Tilde{g}(\bm{\gamma}))^\prime(\sum_{i=1}^n \Tilde{g}(\bm{\gamma}))$.
We represent the minimiser by $\hat{\bm{\gamma}} = (\hat{\gamma}^L, \hat{\gamma}^F, \hat{q}, \hat{\alpha}, \hat{\eta})$.
We can estimate $\rnabla^\mathrm{s}$ by the following:
\begin{align*}
    \widehat{\rnabla^\mathrm{s}_2} = \max_{v \in\{L,F\}} \left(\hat{\gamma}^v - \hat{\eta}\right).
\end{align*}
Now, we have the following lemma:
\begin{lemma}\label{lemma normality under symmetry}
    Under the same assumptions in Corollary \ref{cor normality under symmetry}, $\sqrt{n}\left(\hat{\bm{\gamma}} - \bm{\gamma}_0\right) \to_d \mathcal{N}(0, \Omega_{\gamma,L})$, where $\Omega_{\gamma,L}$ is given in the Online Appendix.
\end{lemma}

Then, we can apply the procedure developed by \cite{Chernozhukov_etal:2013} for inference.
Conceptually, \citeauthor{Chernozhukov_etal:2013}'s (\citeyear{Chernozhukov_etal:2013}) CI is different from \citeauthor{Imbens_Manski:2004}'s (\citeyear{Imbens_Manski:2004}) in that the former is the CI for the identified region, i.e., the bounds $[\rnabla^\mathrm{s}, \nabla^\mathrm{s}]$, while the latter is for the true parameter, $\tau$.
Therefore, the CI based on \cite{Chernozhukov_etal:2013} can be conservative for $\tau$, although valid.

\section{Empirical Illustration}\label{sec: empirical}
\subsection{Empirical Contexts and Main Results}

We illustrate the usefulness of the bounds based on the stochastic monotonicity and symmetry assumptions using the data from \cite{Muralidharan:2019}.
The authors used a randomised controlled trial to examine the effect of the after-school education programme, called Mindspark intervention, on math and Hindi test scores.

The Mindspark intervention consists of two sessions, technology-led and group-based instructions, while the instruction was mainly provided in the formar session (\citealp[p.~1427]{Muralidharan:2019}).
This technology-led instruction is a computer-based interactive instruction. 
The content provided in this session is automatically personalised for each student, and this adaptation is performed dynamically based on the beginning assessment and every subsequent activity completed.

The educational attainment is assessed by the baseline and endline tests. These tests are designed to capture a broad range of students' achievements.
The test questions ranged in difficulty from ``very easy" to ``grade-appropriate" levels (\citealp[p.~1436]{Muralidharan:2019}).

In this intervention, the lottery winners (314 students, $D_i=1$) were assigned to the treatment group, and the losers (305 students, $D_i=0$) were assigned to the control group.
$S_{di}$ denotes whether student $i$ takes the endline test when $D_i=d$.
Note that we define the outcome in interest ($Y_{1i}^*$ and $Y_{0i}^*$) as the difference between the endline and baseline test scores for ease of interpretation, while the authors used a different indicator when computing the \citeauthor{Lee:2009} bounds (\citealp[Table A8]{Muralidharan:2019}).

One problem with the Mindspark intervention was attrition.
$15.3\%$ and $10.5\%$ dropped out from the treatment and control groups, respectively.
To address the potential endogenous sample selection, \cite{Muralidharan:2019} computed \citeauthor{Lee:2009} bounds. They assumed that the monotonicity in the selection, or $S_{0i}\geq S_{1i}$ holds almost surely; this direction is opposite to that in the previous sections.

We first compute the \citeauthor{Lee:2009} bounds based on this \citeauthor{Muralidharan:2019}'s (\citeyear{Muralidharan:2019}) assumption (i.e., $\P{S_{0i}=1 | S_{1i}=1}=1$).
The estimated bounds and \citeauthor{Imbens_Manski:2004}'s (\citeyear{Imbens_Manski:2004}) CIs are reported in rows (1) and (3) of Table \ref{tab: emp 1}.
The CI for the math scores is $(0.153, 0.627)$, and thus, the difference in test scores between the endline and baseline for always-takers is positive.
In contrast, the CI for the Hindi scores is $(-0.019, 0.423)$ and covers zero. This may imply that the possibility of no effect cannot be rejected.

\begin{table}[th]
    \centering
    \scalebox{1}{
    \begin{tabular}{l|cc|cc}
    \hline\hline
         & \multicolumn{2}{c|}{\textbf{Math}} & \multicolumn{2}{c}{\textbf{Hindi}}\\\hline
         & (1) & (2) & (3) & (4)\\
        Symmetry &  & Yes &  & Yes\\
        Bounds & $[0.291, 0.494]$ & $[0.364, 0.424]$ & $[0.110, 0.293]$ & $[0.121, 0.180]$ \\
        CI & $(0.153, 0.627)$ & $(0.227, 0.563)$ & $(-0.019, 0.423)$ & $(0.010, 0.294)$ \\
        \hline
    \end{tabular}}
    \caption{Empirical Bounds under Monotonicity and Symmetry}
    \label{tab: emp 1}
\end{table}

However, this may perhaps be due to the ``extreme" trimming of the \citeauthor{Lee:2009} bounds. To address this point, we estimate bounds under the symmetry assumption, whose validity is discussed in the subsequent subsection.
Now, $1 = \vartheta_L \geq \vartheta_F$ is satisfied, and thus, we can perform the estimation/inference procedure outlined in Section \ref{subsec: known}.
The estimation results are shown in rows (2) and (4) of Table \ref{tab: emp 1}.
The obtained bounds are sharper than the \citeauthor{Lee:2009} bounds. 
The lengths of the bounds for math and Hindi scores are around $70\%$ shorter. 
The CIs are also tightened; their lengths are approximately $29\%$ and $36\%$ shorter than those of the \citeauthor{Lee:2009} bounds.
Consequently, the CI for the Hindi scores does not cover zero, suggesting a positive treatment effect under monotonicity and symmetry.
These results showcase the possible usefulness of the new bounds obtained in Theorem \ref{thm identification under symmetry}.

So far, we have considered bounds when the monotonicity exactly holds.
However, a possible concern is the validity of this monotonicity assumption.
Recall that \cite{Muralidharan:2019} assumed that if a treated student took the end-line test, they must have taken the test if they were in the control group.
This assumption might be justified because the untreated students ``were told that they would be provided free access" to the programme after the end of the experiment if they took the end-line test (\citealp[p.~1434]{Muralidharan:2019}).
However, it would not be surprising if some students with $S_{1i}=1$ dropped out when not in treatment.
For example, students who did not win the lottery might have sought alternative educational resources (e.g., textbooks, other tutoring services) and then might no longer be interested in participating in the Mindspark programme by the time of the endline test. Alternatively, not participating in the Mindspark intervention could cause parents to overlook the endline test, potentially leading to scheduling conflicts (e.g., travel, shopping, assisting with parents' work) on the test day, thereby preventing their children from attending.

Hence, we compute the bounds under stochastic monotonicity. 
Considering the incentive provided to the lottery losers, we assume $\vartheta_L=0.95$. The results are summarised in Table \ref{tab: emp 2}.

\begin{table}[t]
    \centering
    \scalebox{1}{
    \begin{tabular}{l|ccc}
    \hline\hline
         & \multicolumn{3}{c}{\textbf{Math}}\\\hline
         & (1) & (2) & (3)\\
        Symmetry &  & Yes & Yes\\
        Bounds & $[0.222, 0.543]$ & $[0.306, 0.477]$ & $[0.306, 0.477]$\\
        CI & $(0.091, 0.672)$ & $(0.176, 0.610)$ & $(0.129, 0.657)$\\
        \hline
         & \multicolumn{3}{c}{\textbf{Hindi}}\\\hline
         & (4) & (5) & (6)\\
        Symmetry &  & Yes & Yes\\
        Bounds & $[0.048, 0.357]$ & $[0.096, 0.196]$ & $[0.096, 0.196]$\\
        CI & $(-0.075, 0.476)$ & $(-0.010, 0.308)$ & $(-0.048, 0.346)$\\
        \hline
    \end{tabular}}
    \caption{Empirical Bounds under Stochastic Monotonicity ($\vartheta_L = 0.95$) and Symmetry. \textit{Notes: Cases (1), (2), (4), and (5) are based on Section \ref{subsec: known} and \cite{Imbens_Manski:2004}. Cases (3) and (6) are based on Section \ref{subsec: unknown} and \cite{Chernozhukov_etal:2013}.}}
    \label{tab: emp 2}
\end{table}
The results for the math scores are reported in rows (1)--(3).
The treatment effect is still positive even if the monotonicity assumption does not exactly hold, which suggests the robustness of the positive effect of the Mindspark intervention on the math scores.
The results for Hindi are in contrast. Although the bounds and CIs in rows (5)--(6) are much tighter than those without symmetry in row (4), the CIs for the Hindi scores under stochastic monotonicity cover zero. This means that the treatment effect can be zero if we allow the monotonicity to be only slightly violated.
Therefore, the validity of the exact monotonicity will be important to assess whether the Mindspark intervention improved the students' Hindi scores.
This illustrates the importance of relaxing monotonicity and examining the sensitivity of the results under the exact monotonicity.
The stochastic monotonicity assumption is useful for such an objective.

\begin{remark}[Covariates]
    We can utilise an independent covariate to obtain narrower bounds in the same manner as \citet[p.~1086]{Lee:2009}. 
    However, in our empirical example, the qualitative implication of the bounds (e.g., whether the bounds and CIs include zero or not) is unchanged for all cases even if we use such a covariate.
    Hence, these results are postponed to the Online Appendix. Some technical arguments are also provided.
\end{remark}

\subsection{Validity of Assumptions}\label{subsec: validity}
Below, we provide an illustrative discussion of the validity of our assumptions.
\subsubsection*{Mean-Median Coincidence}
In the context of the Mindspark intervention, the mean-median coincidence assumption could be reasonable.
As noted before, the main instruction provided via Mindspark software is automatically personalised. Besides, the tests include a wide range of questions from easy to difficult, designed to measure the various improvements of students.
Considering these points, we may assume that the improvement in test scores is similar across students, and the observed differences may stem from stochastic reasons, such as compatibility with the questions. Then, we may assume $Y_{1i}^* = t + \varepsilon_i$, where $t\in\mathbb{R}$ and $\E{\varepsilon_i} = \mathrm{Median}[\varepsilon_i] = 0$ such as the Gaussian error, which implies Assumption \ref{assumption mean median}.
This certain homogeneity assumption is consistent with \citet[Section III.C]{Muralidharan:2019}, wherein they found limited evidence of heterogeneity in students' progress by their initial learning level, gender, and socioeconomic status.

\subsubsection*{Tail Smoothness}
\begin{figure}[ht]
    \centering
    \includegraphics[scale=0.5]{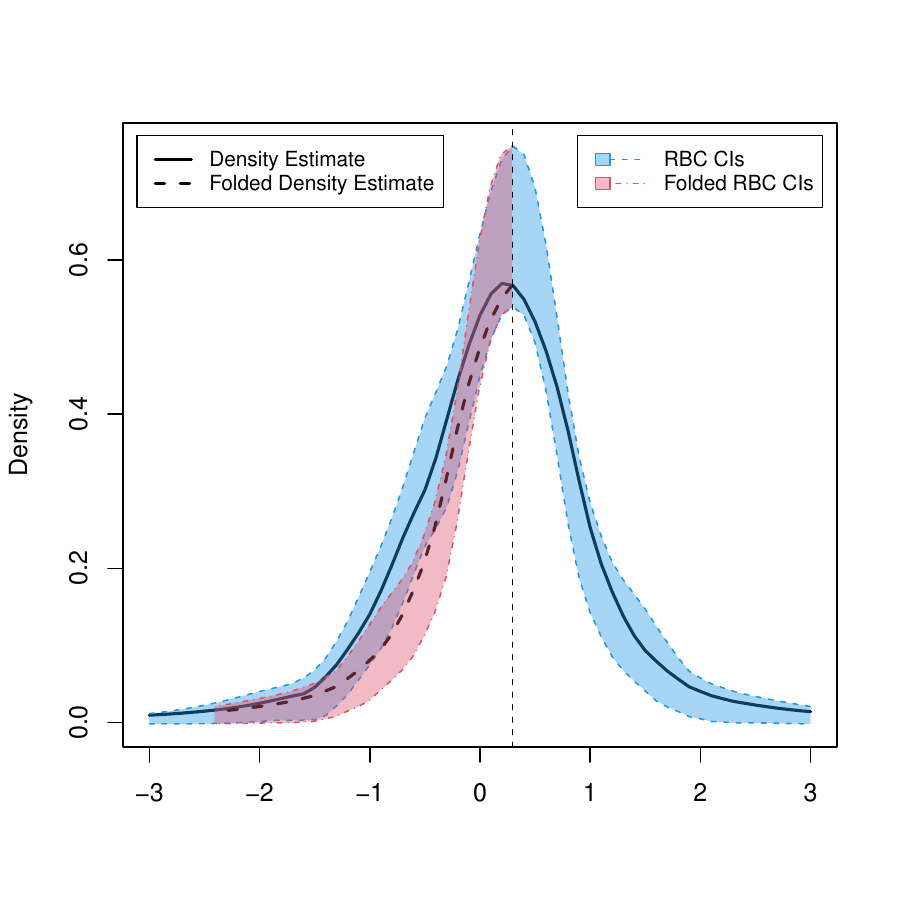}
    \caption{Density Estimate of $f_y^0$ (Hindi)}
    \label{fig: KDE sharpness}
\end{figure}

We assess the sharpness of the lower bound on Hindi scores under symmetry (Table \ref{tab: emp 2} (5)--(6)), which may be the most intriguing aspect, by visually verifying Assumption \ref{assumption: tail}.
Let $F_0$ and $f_y^0$ be the distribution and density functions, respectively, of $Y_i$ conditional on $D_i=0$ and $S_i=1$; and $y^0_r = F_0^{-1}(r)$.
In Figure \ref{fig: KDE sharpness}, we show the kernel density estimates of $f_y^0$, 99\% point-wise robust bias-corrected (RBC) CIs of \cite{calonico2018effect}, and their folded ones at $\hat{y}^0_{1-\hat{\vartheta}\hat{q}_0 /2}$, which is indicated by a dotted vertical line. We find no strong evidence suggesting Assumption \ref{assumption: tail} is refuted; the folded line is always below the original density estimates, and CIs do not indicate that the former is above the latter. Hence, the sharpness may not be rejected.

\subsubsection*{Stochastic Monotonicity}
When the upper bound in \eqref{P upper} is too small, it may suggest the implausibility of Assumption \ref{assumption: stochastic monotonicity}-(i).
As the direction of the monotonicity is opposite in our example, we compute the empirical upper bound on $\P{S_{0i} = 1 | S_{1i} = 0}$, finding it to be $0.56$, which is not unreasonably small.
Assumption \ref{assumption: stochastic monotonicity}-(ii) may be justified by the premise that most of the students with $S_{1i}=1$ are motivated and would likely take the endline test to participate in the program if they lose the lottery, and that the attrition resulting from the previously mentioned taste shock or overlapping schedules can be regarded as a random shock.

\section{Conclusion}\label{sec: conclusion}
This study considered the partial identification of \citeauthor{Lee:2009}'s general sample selection model under a stochastic version of the monotonicity and symmetry assumptions.
The former assumption was introduced to robustify the \citeauthor{Lee:2009} bounds to the small deviation from the exact monotonicity, while the latter assumption can be used to tighten the \citeauthor{Lee:2009} bounds, which may be empirically wide.
The obtained bounds are root-$n$ consistently estimable, making it easy to perform estimation and inference in practice.

Extensive literature has applied the results of \cite{Lee:2009} or \cite{Horowitz_Manski:1995}. 
Our proposed stochastic monotonicity assumption can be applied in many such contexts. 
As an important example, we treat \cite{Bartalotti_etal:2023} in the Online Appendix.
Furthermore, the simplicity of such an extension will justify theoretical studies based on the monotonicity assumption.

\bibliographystyle{apalike}
\bibliography{refs}

\section*{Appendix: Proof of Theorems}
\renewcommand{\theequation}{A.\arabic{equation}}
\renewcommand{\thesection}{A}
\setcounter{equation}{0}

\medskip
In the proof of Theorem \ref{thm identification}, we utilise Proposition 4 of \cite{Horowitz_Manski:1995}. For the reader's convenience, we reproduce their statement below. 
Let $F$ and $G$ are distributions and $F$ (first-order) stochastically dominates $G$.
We say that a $\mathbb{R}$-valued functional $\rho$ respects stochastic dominance if $\rho(F) \geq \rho(G)$.\\

\noindent
\textbf{Proposition 4 of \cite{Horowitz_Manski:1995}:}
    Let $Y$ be a continuous random variable from a distribution $Q$ such that $Q = pP_0 + (1-p)P_1$, and $y_q$ be the $q$-th quantile of $Q$. Assume we know that $p\leq\lambda<1$ and $\rho$ respects stochastic dominance. Define the distributions
    \begin{align*}
        L_\lambda(t) &= Q(t)/(1-\lambda)\times\mathbf{1}\{t < y_{1-\lambda}\} + \mathbf{1}\{t \geq y_{1-\lambda}\},\\
        U_\lambda(t) &= \left(Q(t)-\lambda\right)/(1-\lambda)\times\mathbf{1}\{t \geq y_{\lambda}\}.
    \end{align*}
    Then, $\rho(P_1) \in \left[\rho(L_\lambda), \rho(U_\lambda)\right]$ and the bounds are sharp.\\

\noindent
\textbf{Proof of Theorem \ref{thm identification}:}
Note that 
\begin{align*}
    \E{Y_{0i}^*|S_{0i}=S_{1i}=1} = \E{Y_{0i}^*|S_{0i}=1} = \E{Y_{i}|D_i=0,S_i=1},
\end{align*}
where the first equality follows from Assumption \ref{assumption: stochastic monotonicity}-(ii) and the second from Assumption \ref{assumption: independence}.
Thus, it suffices to show the first terms of $\nabla$ and $\rnabla$ are sharp bounds.
Below, we assume $\vartheta_L<1$ as the case $\vartheta_L=1$ reduces to the \citeauthor{Lee:2009}'s (\citeyear{Lee:2009}) result.
Observe
\begin{align}
    &\P{Y_i\leq y|D_i=1, S_i=1}\notag\\
    &\quad=
     \P{Y_{1i}^*\leq y|S_{0i}=0, S_{1i}=1} p_0
     +\P{Y_{1i}^*\leq y|S_{0i}=1, S_{1i}=1} (1-p_0),\label{decomposition}
\end{align}
where $p_0=\P{S_{0i}=0,S_{1i}=1|D_i=1}/\P{S_{i}=1|D_i=1}$.
By Assumption \ref{assumption: independence} and Assumption \ref{assumption: stochastic monotonicity}, $1-p_0$ can be bounded from below as
\begin{align*}
    1-p_0&=\frac{\P{S_{0i}=1,S_{1i}=1|D_i=1}}{\P{S_{i}=1|D_i=1}}
    = \frac{\P{S_{0i}=1,S_{1i}=1|D_i=0}}{\P{S_{i}=1|D_i=1}}\\
    &= \frac{\P{S_{i}=1|D_i=0}\P{S_{1i}=1|S_{0i}=1}}{\P{S_{i}=1|D_i=1}}
    \geq \vartheta_L q_0,
\end{align*}
or equivalently, $p_0\leq 1-\vartheta_L q_0$.
Plus, applying the Fréchet inequality to $\P{S_{1i}=1|S_{0i}=1}$, we have $1-p_0 \geq \vartheta_F q_0$, i.e.,  $p_0 \leq 1- \vartheta_F q_0$. 
Hence, $p_0\leq 1-\vartheta q_0$ with $\vartheta=\max\{\vartheta_L, \vartheta_F\}$. Note that the expectation operator respects stochastic dominance. Then, applying Proposition 4 of \cite{Horowitz_Manski:1995} to (\ref{decomposition}), we find that the first terms of $\nabla$ and $\rnabla$ are the sharp bounds on $\E{Y_{1i}^* | S_{01}= S_{1i}=1}$. 
\hfill$\square$\\

\noindent
\textbf{Proof of Theorem \ref{thm identification under symmetry}:}
We focus on the upper bound as the proof for the lower bound is similar.
Firstly, we show the validity of $\nabla^{\mathrm{s}}$, that is, 
\begin{align*}
    \E{Y_{1i}^*|S_{0i}=S_{1i}=1} \leq y^1_{1 - \frac{\vartheta q_0}{2}}.
\end{align*}
To show this, assume $\E{Y_{1i}^*|S_{0i}=S_{1i}=1} > y^1_{1 - {\vartheta q_0}/{2}}$.
Recalling \eqref{decomposition}, it follows that
\begin{align*}
    \P{Y_i > y^1_{1 - {\vartheta q_0}/{2}}|D_i=1, S_i=1}
    \geq
    \P{Y_{1i}^* > y^1_{1 - {\vartheta q_0}/{2}} |S_{0i}=1, S_{1i}=1} (1-p_0).
\end{align*}
Then, we can compute that
\begin{align}
    \frac{\vartheta q_0}{2} &\geq 
    \P{Y_i > y^1_{1 - {\vartheta q_0}/{2}}|D_i=1, S_i=1}\notag\\
    &\geq
    \P{Y_{1i}^* > y^1_{1 - {\vartheta q_0}/{2}} |S_{0i}=1, S_{1i}=1} (1-p_0)\notag\\
    &>
    \frac{1}{2} \times (1-p_0)\label{1/2 by symmetry}\\
    &\geq \frac{\vartheta q_0}{2},\notag
\end{align}
where the first inequality uses the definition of $y_r^1$, and \eqref{1/2 by symmetry} holds under Assumption \ref{assumption mean median} and $\E{Y_{1i}^*|S_{0i}=S_{1i}=1} > y^1_{1 - {\vartheta q_0}/{2}}$. Thus, contradiction arrises and $\E{Y_{1i}^*|S_{0i}=S_{1i}=1} \leq y^1_{1 - {\vartheta q_0}/{2}}$ is proved. 
The validity of the bounds under Assumption \ref{assumption: symmetry} is now obvious as Assumption \ref{assumption: symmetry} is sufficient for Assumption \ref{assumption mean median}.

Next, we additionally assume Assumption \ref{assumption: tail}.
We will show that the upper bound $\nabla^{\mathrm{s}}$ is attainable.
We construct a joint distribution of $(\Tilde{Y}_{1i}^*, \Tilde{Y}_{0i}^*, \Tilde{S}_{1i}, \Tilde{S}_{0i}, D_i)$ that is consistent with the assumptions, attains the upper bound, and induces the joint distribution of $(Y_i, S_i, D_i)$.
Firstly, define
\begin{align*}
    &\P{\Tilde{S}_{0i} = 1} \coloneqq \P{S_{i}=1 | D_i=0}, \, 
    \P{\Tilde{S}_{1i} = 1} \coloneqq \P{S_{i}=1 | D_i=1}, \\
    &\P{\Tilde{Y}_{0i}^* \leq y \big| \Tilde{S}_{0i} = 1} \coloneqq \P{Y_i \leq y | D_i=0, S_i=1}, \\
    &\P{\Tilde{Y}_{1i}^* \leq y \big| \Tilde{S}_{1i} = 1} \coloneqq \P{Y_i \leq y | D_i=1, S_i=1},
\end{align*}
and $\P{\Tilde{S}_{1i}=1 | \Tilde{S}_{0i}=1} \coloneqq \vartheta$.
Then, we can obtain that
\begin{align*}
    \pi_{11}\coloneqq &\P{\Tilde{S}_{0i}=1, \Tilde{S}_{1i}=1} = \vartheta\P{S_{i}=1 | D_i=0},\\
    \pi_{10}\coloneqq &\P{\Tilde{S}_{0i}=1, \Tilde{S}_{1i}=0} = (1-\vartheta)\P{S_{i}=1 | D_i=0},\\
    \pi_{01}\coloneqq &\P{\Tilde{S}_{0i}=0, \Tilde{S}_{1i}=1} = \P{S_{i}=1 | D_i=1} - \vartheta\P{S_{i}=1 | D_i=0},\\
    \pi_{00}\coloneqq &\P{\Tilde{S}_{0i}=0, \Tilde{S}_{1i}=0} = 1 - \pi_{11} - \pi_{10} - \pi_{01},
\end{align*}
and $\P{\Tilde{Y}_{di}^* \leq y, \Tilde{S}_{di} = 1} = \P{Y_i \leq y , S_i=1 | D_i=d}$, $d\in\{0,1\}$.
Next, we define
\begin{align}
    &\pi^*_{1|1}\coloneqq\P{\Tilde{Y}_{1i}^* \leq y \big| \Tilde{S}_{0i} = 1, \Tilde{S}_{1i} = 1} \notag\\
    &\qquad\qquad \coloneqq
    \begin{cases}
        \dfrac{1}{2} + \dfrac{1}{\vartheta q_0} \P{y^1_{1-\frac{\vartheta q_0}{2}} \leq \Tilde{Y}_{1i}^* \leq y \big| \Tilde{S}_{1i}=1 }& \text{if } y \geq y^1_{1-\frac{\vartheta q_0}{2}}\\
        \dfrac{1}{\vartheta q_0} \P{2 y^1_{1-\frac{\vartheta q_0}{2}} - y \leq \Tilde{Y}_{1i}^* \big| \Tilde{S}_{1i}=1 } & \text{if } y < y^1_{1-\frac{\vartheta q_0}{2}}
    \end{cases},\notag\\
    &\P{\Tilde{Y}_1^* \leq y \big| \Tilde{S}_{0i} = 1, \Tilde{S}_{1i} = 0} \coloneqq \pi^*_{1|1},\notag\\
    &\P{\Tilde{Y}_1^* \leq y \big| \Tilde{S}_{0i} = 0, \Tilde{S}_{1i} = 1} \coloneqq \frac{1}{1-\vartheta q_0}\P{\Tilde{Y}_{1i}^* \leq y | \Tilde{S}_{1i} = 1} - 
    \frac{\vartheta q_0}{1-\vartheta q_0}\pi^*_{1|1},\label{conditional prob}\\
    &\P{\Tilde{Y}_1^* \leq y \big| \Tilde{S}_{0i} = 0, \Tilde{S}_{1i} = 0} \coloneqq \P{\Tilde{Y}_1^* \leq y \big| \Tilde{S}_{1i} = 1}\notag
\end{align}
and
\begin{align*}
    &\P{\Tilde{Y}_0^* \leq y \big| \Tilde{S}_{0i} = s_0, \Tilde{S}_{1i} = s_1} \coloneqq \P{\Tilde{Y}_0^* \leq y \big| \Tilde{S}_{0i} = 1}
\end{align*}
for all $s_0,s_1\in\{0,1\}$.
Note that Assumptions \ref{assumption: stochastic monotonicity} and \ref{assumption: symmetry} are satisfied. Also, \eqref{conditional prob} is a proper distribution function under Assumption \ref{assumption: tail}.
Assume $\Tilde{Y}_{1i}^* \indep \Tilde{Y}_{0i}^* | (\Tilde{S}_{1i},\Tilde{S}_{0i})$, then we can write the joint distribution as a product of the defined conditional probabilities under Assumption \ref{assumption: independence}. Besides, letting 
\begin{align*}
& \Tilde{S}_i=\Tilde{S}_{1i}D_i + \Tilde{S}_{0i}(1-D_i),\,
\Tilde{Y}_i = \Tilde{S}_i\cdot\left\{\Tilde{Y}_{1i}^* D_i + \Tilde{Y}_{0i}^* (1-D_i)\right\},
\end{align*}
it follows that
\begin{align*}
    \P{\Tilde{Y}_i \leq y, \Tilde{S}_i=1, D_i=1} =
    \P{\Tilde{Y}_{1i}^* \leq y, \Tilde{S}_{1i}=1}\P{D_i=1} =
    \P{Y_i \leq y , S_i=1 , D_i=1}
\end{align*}
and that $\P{\Tilde{Y}_i \leq y, \Tilde{S}_i=1, D_i=0} = \P{Y_i \leq y , S_i=1 , D_i=0}$ similarly. Hence, it induces the joint distribution of the data $(Y_i, S_i, D_i)$. 
Finally,
\begin{align*}
    \nabla^{\mathrm{s}} &= 
    \E{\Tilde{Y}_1^* \big| \Tilde{S}_{0i} = 1, \Tilde{S}_{1i} = 1} - 
    \E{\Tilde{Y}_0^* \big| \Tilde{S}_{0i} = 1}
    = \E{\Tilde{Y}_1^* - \Tilde{Y}_0^* \big| \Tilde{S}_{0i} = 1, \Tilde{S}_{1i} = 1},
\end{align*}
where the first equality follows by construction. This shows that the upper bound is attainable.
\hfill$\square$\\

\newpage
\setcounter{page}{1}
\renewcommand{\thepage}{S\arabic{page}}
\setcounter{equation}{0}
\renewcommand{\theequation}{S\arabic{equation}}
\setcounter{section}{0}
\renewcommand{\lemma}{S\arabic{lemma}}
\setcounter{theoremx}{0}
\renewcommand{\thetheoremx}{S\arabic{theoremx}}
\setcounter{table}{0}
\renewcommand{\thesection}{S\arabic{section}}
\setcounter{table}{0}
\renewcommand{\thetable}{S\arabic{table}}
\setcounter{figure}{0}
\renewcommand{\thefigure}{S\arabic{figure}}
\setcounter{remarkx}{0}
\renewcommand{\theremarkx}{S\arabic{remarkx}}
\setcounter{exx}{0}
\renewcommand{\theexx}{S\arabic{exx}}


\makeatletter
\renewcommand*{\@fnsymbol}[1]{\ensuremath{\ifcase#1\or \flat\or * \else\@ctrerr\fi}}
\makeatother

\begin{center}
{\Large\bf Online Appendix for\\ ``Robustify and Tighten the Lee Bounds: \\ A Sample Selection Model under Stochastic Monotonicity and Symmetry Assumptions"}\\
\vspace{0.5cm}
{\large Yuta Okamoto$^\sharp$}\\
\vspace{0.3cm}
{\large $^\sharp$Graduate School of Economics, Kyoto University}\\
\vspace{0.5cm}
{March 20, 2024}\\
\vspace{1cm}
\end{center}

\section{Omitted Results and Proofs}

\subsection{Proofs}
\textbf{Proof of Corollary 1:}
This is an immediate consequence of Theorem 3.1.\hfill$\square$\\

\noindent
\textbf{Proof of Corollary 2:}
Assuming the parameter space for $q$ be $[0,1/\vartheta_L]$, the consistency, $\hat{\bm{\beta}}\to_p \bm{\beta}_0$, holds under the boundedness assumption of the support of $Y_i$ and $\P{S_i=1 | D_i=0}>0$.
This follows from a similar argument to Proposition 2 of \citet{Lee:2009}, utilising Theorem 2.6 from \cite{Newey_McFadden:1994}.
Define $g_0(\bm{\beta}) = \E{g(\bm{\beta})}$ and $\hat{g}_n(\bm{\beta}) = n^{-1} \sum_{i=1}^n g(\bm{\beta})$.
Let $G_\beta$ denote the derivative of $g_0$ at $\bm{\beta}=\bm{\beta}_0$ and $\Sigma_\beta$ be the asymptotic variance of $\hat{g}_n(\bm{\beta}_0)$.
By \citet[Theorem 7.2]{Newey_McFadden:1994} and \citet[Theorems 1--3]{Andrews:1994}, under the stated assumptions, we have the asymptotic normality of $\hat{\bm{\beta}}$ with the asymptotic variance $G_\beta^{-1}\Sigma_\beta (G_\beta^\prime)^{-1}$. Here, we can compute 
\begin{align*}
    G_\beta &= \begin{pmatrix}
        -f^1_y\left(y^1_{\frac{\vartheta_L q_0}{2}}\right)\E{S_i D_i} & \dfrac{\vartheta_L}{2}\E{S_i D_i} & 0 & 0\\
        0 & \E{S_i D_i} & -\P{D_i=1} & 0\\
        0 & 0 & -\P{D_i=0} & 0\\
        0 & 0 & 0 & -\E{S_i (1 - D_i)}
    \end{pmatrix},
\end{align*}
and $\Sigma_\beta = \mathrm{diag}(\Sigma_{1}, \Sigma_3, \Sigma_4, \Sigma_5)$, where these elements are defined in the proof of Lemma 1 below.
A straightforward calculation yields that
\begin{align*}
    G_{\beta}^{-1} &= \left\{f^1_y\left(y^1_{\frac{\vartheta_L q_0}{2}}\right)\right\}^{-1}\dfrac{1}{\P{S_i=1,D_i=1}}\begin{pmatrix}
        -1 & \dfrac{\vartheta_L}{2} & 
        -\dfrac{\vartheta_L}{2}\dfrac{\P{D_i=1}}{\P{D_i=0}} & 0\\
        0 & C_{1} & C_{2} & 0\\
        0 & 0 & C_{3} & 0\\
        0 & 0 & 0 & 0
    \end{pmatrix} + M_\eta^{-1},
\end{align*}
where $C_{j}$ are some constants and $M_\eta^{-1}$ is a $4\times4$ matrix where the bottom right element is $-1/\E{S_i(1-D_i)}$ and all others are 0.
Then, the asymptotic variance of $\hat{\beta}^L - \hat{\eta}$ is given by 
\begin{align*}
    \left\{f^1_y\left(y^1_{\frac{\vartheta_L q_0}{2}}\right)^2\right\}^{-1}
    \frac{1}{\P{S_i=1,D_i=1}^2}
    \left\{
    \Sigma_{1} + \frac{\vartheta_L^2}{4}\Sigma_3 + \frac{\vartheta_L^2}{4}\frac{\P{D_i=1}^2}{\P{D_i=0}^2}\Sigma_4
    \right\} + \Omega_C.
\end{align*}
The asymptotic variance of the upper bound is similar.\hfill$\square$\\

\noindent
\textbf{Proof of Lemma 1:}
The asymptotic variance is given by $\Omega_{\gamma,L} = G_{\gamma,L}^{-1}\Sigma_{\gamma}(G_{\gamma,L}^{\prime})^{-1}$; $G_{\gamma,L}$ is given by
\begin{align*}
    \begin{pmatrix}
        -f_y^1(\gamma^L_0) \E{S_i D_i} & 0 & \dfrac{\vartheta_L}{2} \E{S_i D_i} & 0 & 0\\
        0 & -f_y^1(\gamma^F_0) \E{S_i D_i} & \dfrac{1-1/\alpha_0}{2}\E{S_i D_i} & \dfrac{q_0}{2 \alpha_0^2}\E{S_i D_i} & 0\\
        0 & 0 & \E{S_i D_i} & -\E{D_i} & 0\\
        0 & 0 & 0 & -\E{1 - D_i} & 0\\
        0 & 0 & 0 & 0 & 0\\
    \end{pmatrix} + M_\eta;
\end{align*}
$M_\eta$ is a $5\times5$ matrix where the bottom right element is $-\E{S_i(1-D_i)}$ and all others are 0; and $\Sigma_{\gamma} = \mathrm{diag}(\Sigma_1, \Sigma_2, \Sigma_3, \Sigma_4, \Sigma_5)$, where
\begin{align*}
    \Sigma_{1} &= 
    \frac{\vartheta_L q_0}{2}\left(1 - \frac{\vartheta_L q_0}{2}\right)\P{S_i=1,D_i=1},\,
    \Sigma_{2} = 
    \frac{\vartheta_F q_0}{2}\left(1 - \frac{\vartheta_F q_0}{2}\right)\P{S_i=1,D_i=1},\\
    \Sigma_3 &= {\alpha_0}\left(q_0 - {\alpha_0}\right)\P{D_i=1},\,
    \Sigma_4 = \alpha_0(1-\alpha_0)\P{D_i=0},
\end{align*}
and $\Sigma_5=\V{Y_i | S_i=1, D_i=0}\E{S_i(1-D_i)}$.
The proof is similar to that of Corollary 2 above.\hfill$\square$\\

\subsection{Estimation and Inference without Symmetry}
Assume a researcher knows that $\vartheta_L \geq \vartheta_F$ but hesitates to impose the symmetry. In such a case, one can estimate the lower bound based on 
\begin{align*}
    h(\bm{\beta}) = 
    \begin{pmatrix}
        (Y_i - \mu) S_i D_i \mathbf{1}\{Y_i \leq \beta^L\}\\
        \left(\mathbf{1}\left\{Y_i > \beta^L\right\} - \left(1-\vartheta_L q\right)\right) S_i D_i\\
        \left(q S_i - \alpha\right) D_i\\
        (S_i - \alpha)(1 - D_i)\\
        (Y_i - \eta) S_i (1 - D_i)
    \end{pmatrix}
\end{align*}
with an additional parameter $\mu$ and $\mu_0 = \E{Y_i|D_i=1, S_i = 1, Y_i\leq y^1_{\vartheta_L q_0}}$, instead of $g(\bm{\beta})$.
The estimator is given by $\widehat{\rnabla}_1 = \hat{\mu} - \hat{\eta}$. $\widehat{\nabla}_1$ is defined similarly.
\begin{corollary}
Under the same assumptions in Corollary 2, $\sqrt{n}\left(\widehat{\rnabla}_1 - \rnabla\right) \to_d \mathcal{N}(0, \Tilde{\Omega}_L + \Omega_C)$ and $\sqrt{n}\left(\widehat{\nabla}_1 - \nabla\right) \to_d \mathcal{N}(0, \Tilde{\Omega}_U + \Omega_C)$, where
\begin{align*}
     \Tilde{\Omega}_{L} &= \frac{\V{Y_i | D_i=1, S_i=1, Y_i\leq y_{\vartheta_L q_0}^1} + 
    \left(y_{\vartheta_L q_0}^1 - \mu_L\right)^2 (1-\vartheta_L q_0)}{\P{S_i=1, D_i=1}\vartheta_L q_0} +
    \left({y_{\vartheta_L q_0}^1 - \mu_L}\right)^2 \Omega_{Q},\\
    \Tilde{\Omega}_{U} &= \frac{\V{Y_i | D_i=1, S_i=1, Y_i\geq y_{1-\vartheta_L q_0}^1} + 
    \left(y_{1-\vartheta_L q_0}^1 - \mu_U\right)^2 (1-\vartheta_L q_0)}{\P{S_i=1, D_i=1}\vartheta_L q_0} +
    \left({y_{1-\vartheta_L q_0}^1 - \mu_U}\right)^2 \Omega_{Q},\\
    \Omega_{Q} &= \frac{1-{\alpha_0}/{q_0}}{\P{D_i=1}{\alpha_0}/{q_0} 
    }+ \frac{1-\alpha_0}{\alpha_0\P{D_i=0}},
 \end{align*}
 $\mu_L = \E{Y_i|D_i=1, S_i = 1, Y_i\leq y^1_{\vartheta_L q_0}}$, $\mu_U = \E{Y_i|D_i=1, S_i = 1, Y_i\geq y^1_{1-\vartheta_L q_0}}$,
 and $\Omega_C$ is defined in Corollary 2.
 \end{corollary}
\begin{remark}
    The asymptotic variances with $\vartheta_L=1$ coincide with those in \citet[Proposition 3]{Lee:2009}.
\end{remark}

When a researcher does not know that $\vartheta_L \geq \vartheta_F$ and hesitates to impose the symmetry.
We shall replace $\Tilde{g}$ with 
\begin{align*}
    \Tilde{h}(\bm{\gamma}) = 
    \begin{pmatrix}
        (Y_i - \mu^L) S_i D_i \mathbf{1}\{Y_i \leq \gamma^L\}\\
        (Y_i - \mu^F) S_i D_i \mathbf{1}\{Y_i \leq \gamma^F\}\\
        \left(\mathbf{1}\left\{Y_i > \gamma^L\right\} - \left(1-\vartheta_L q\right)\right) S_i D_i\\
        \left(\mathbf{1}\left\{Y_i > \gamma^F\right\} - \left(1-(1 + q(1-1/\alpha))\right)\right) S_i D_i\\
        \left(q S_i - \alpha\right) D_i\\
        (S_i - \alpha)(1 - D_i)\\
        (Y_i - \eta) S_i (1 - D_i)
    \end{pmatrix},
\end{align*}
where $\mu^L, \mu^F$ are additional parameters and
\begin{align*}
    \widehat{\rnabla_2} = \max_{v \in\{L,F\}} \left(\hat{\mu}^v - \hat{\eta}\right).
\end{align*}
Inference can be performed similarly to that we considered in the main text based on \cite{Chernozhukov_etal:2013}.

\section{Empirical Illustration: Additional Results}
We can use independent covariates $W_i$ in the same manner as \citet[p.~1086]{Lee:2009} to narrow bounds in Theorems 1 and 2. Formally, the covariates satisfying
\begin{align*}
    (Y_{1i}^*, Y_{0i}^*, S_{1i}, S_{0i}, W_i) \indep D_i\text{ and }
    W_i \indep S_{1i} | S_{0i}=1
\end{align*}
can be used. With this $W_i$, we first obtain the lower bound for each $W_i=w$, written by $\rnabla(w)$, and then we obtain the narrower bounds by computing
\begin{align*}
    \int \rnabla(w) f_w (w | S_{i} = 1, D_i=0)\, dw,
\end{align*}
where $f_w$ is the conditional density of the covariates. The upper bound is similar.
    
In Table \ref{tab: emp w/ cov}, we report the covariate-adjusted bounds.
Following \citet[Table A8]{Muralidharan:2019}, we use the location where the Mindspark intervention was held (Mindspark centres) as a covariate.
\begin{table}[ht]
    \centering
    \scalebox{1}{
    \begin{tabular}{l|cc|cc}
    \hline\hline
         \multicolumn{5}{l}{\textbf{Bounds under Monotonicity ($\vartheta=1$)}}\\\hline
         & \multicolumn{2}{c|}{\textbf{Math}} & \multicolumn{2}{c}{\textbf{Hindi}}\\\hline
         & (1) & (2) & (3) & (4)\\
        Symmetry &  & Yes &  & Yes \\
        Bounds & $[0.320, 0.497]$ & $[0.356, 0.465]$ & $[0.084, 0.252]$ & $[0.151, 0.247]$ \\
        CI & $(0.172, 0.632)$ & $(0.225, 0.598)$ & $(-0.038, 0.409)$ & $(0.031, 0.374)$  \\\hline
        \multicolumn{5}{l}{\textbf{Bounds under Stochastic Monotonicity ($\vartheta=0.95$)}}\\\hline
         & \multicolumn{2}{c|}{\textbf{Math}} & \multicolumn{2}{c}{\textbf{Hindi}}\\\hline
         & (1) & (2) & (3) & (4)\\
        Symmetry &  & Yes &  & Yes \\
        Bounds & $[0.247, 0.560]$ & $[0.318, 0.496]$ & $[0.076, 0.345]$ & $[0.107, 0.242]$ \\
        CI & $(0.114, 0.687)$ & $(0.189, 0.628)$ & $(-0.049, 0.469)$ & $(-0.010, 0.364)$ \\\hline
    \end{tabular}}
    \caption{Empirical Bounds with Covariate}
    \label{tab: emp w/ cov}
\end{table}

\section{Log Wage Distributions of American Workers}
\begin{figure}[ht]
    \centering
    \begin{subfigure}[b]{0.49\textwidth}
        \centering
        \includegraphics[scale=0.4]{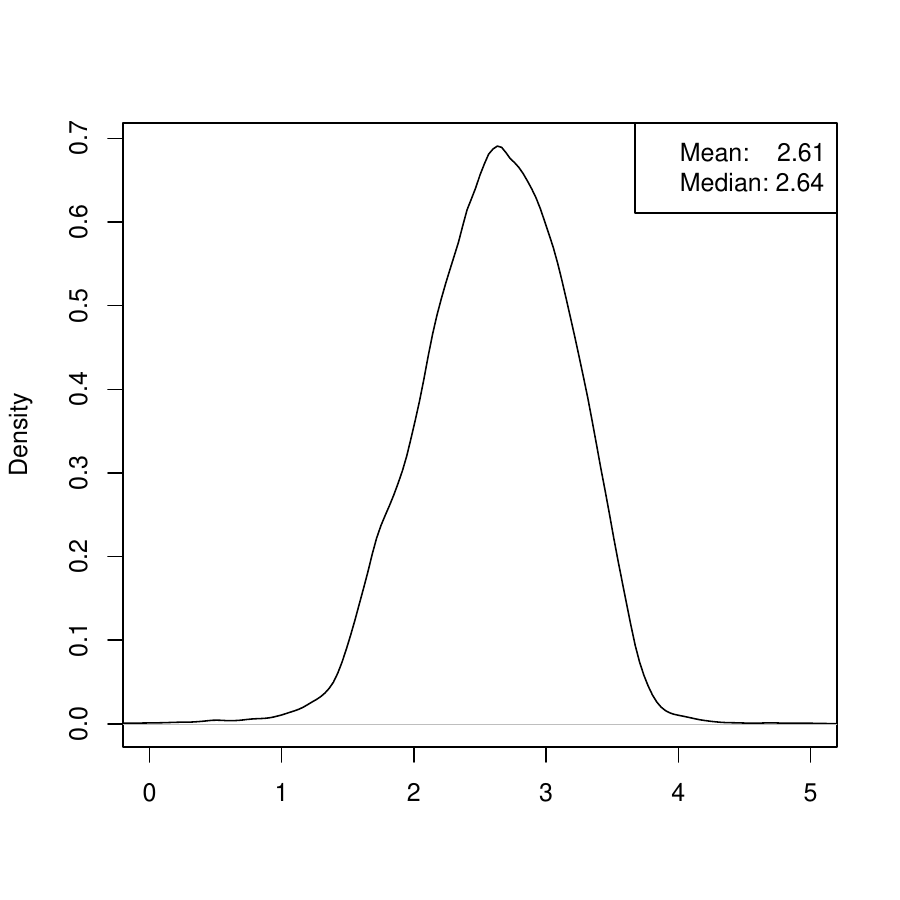}
        \caption{White Men}
        \label{fig: wage1}
    \end{subfigure}
    \hfill
    \begin{subfigure}[b]{0.49\textwidth}
        \centering
        \includegraphics[scale=0.4]{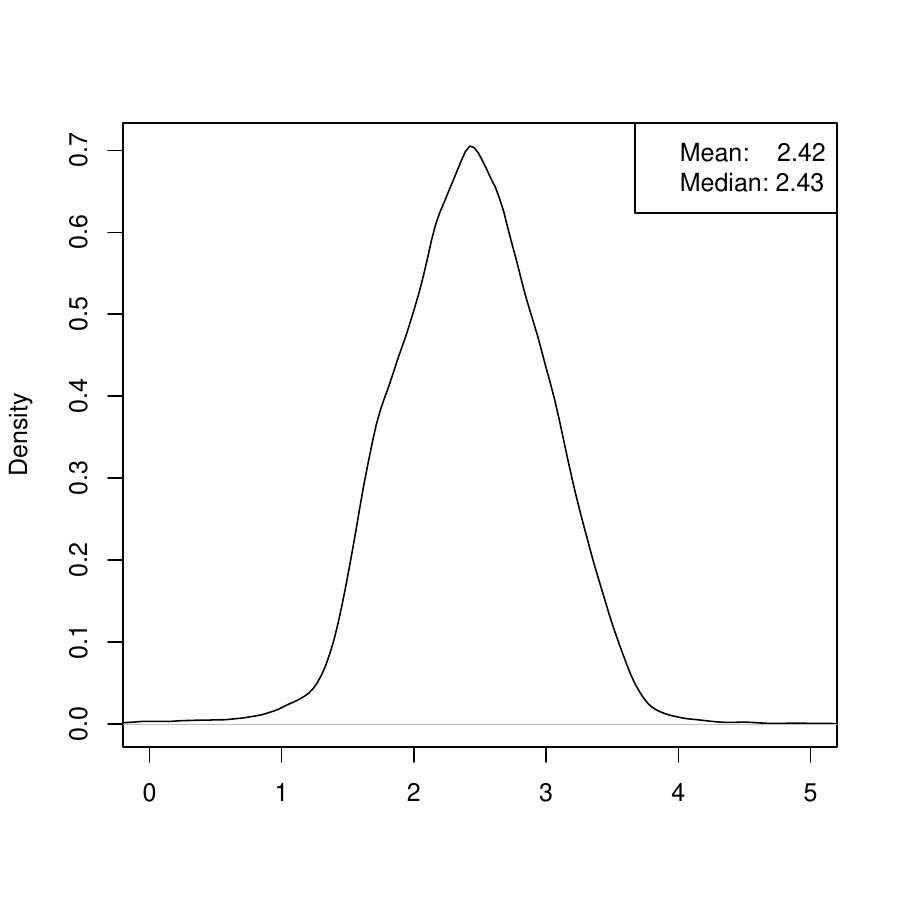}
        \caption{White Women}
        \label{fig: wage2}
    \end{subfigure}
    \\
    \begin{subfigure}[b]{0.49\textwidth}
        \centering
        \includegraphics[scale=0.4]{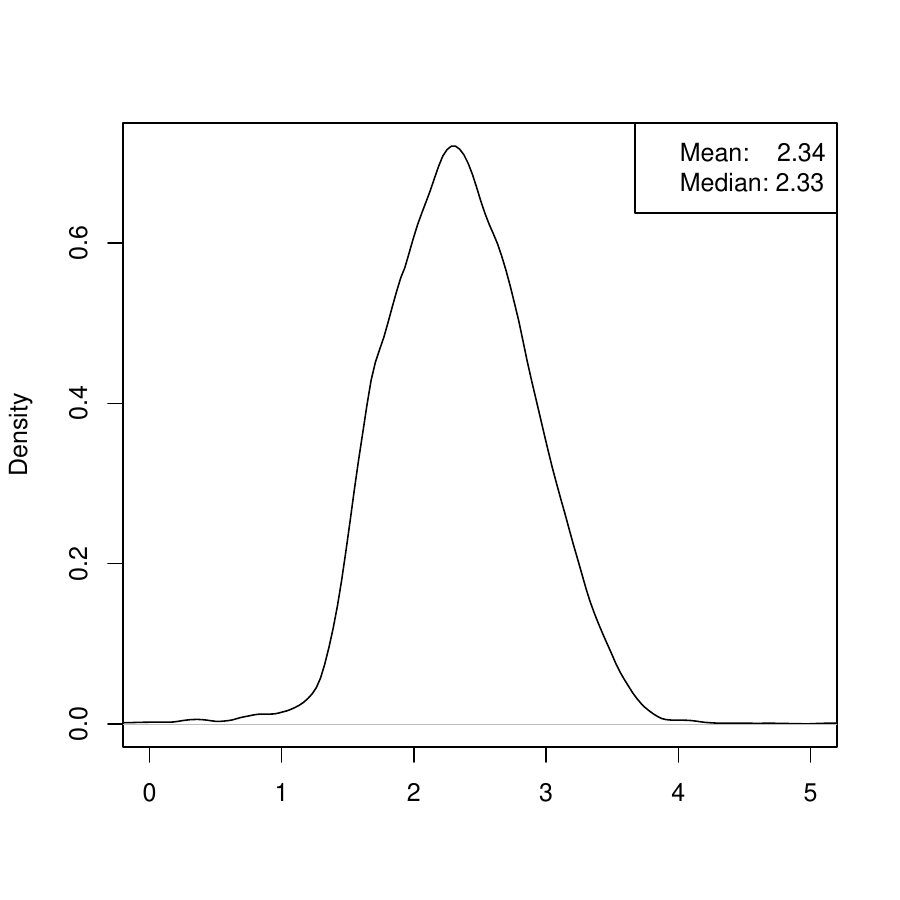}
        \caption{Mexican Men}
        \label{fig: wage3}
    \end{subfigure}
    \hfill
    \begin{subfigure}[b]{0.49\textwidth}
        \centering
        \includegraphics[scale=0.4]{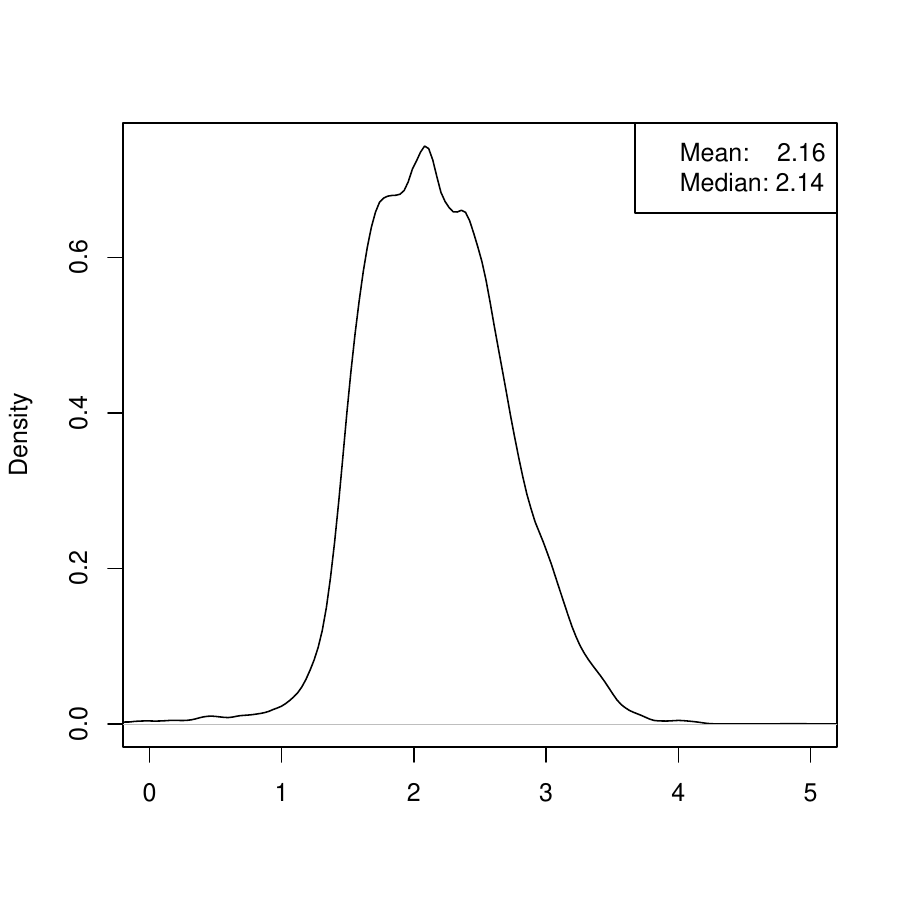}
        \caption{Mexican Women}
        \label{fig: wage4}
    \end{subfigure}
    \caption{Log Wage Distributions}
\end{figure}

We present density estimates for the log wages of American workers. The dataset is sourced from \cite{Honore_Hu:2020}. Following \cite{Honore_Hu:2020}, our analysis is confined to samples from Arizona, California, New Mexico, and Texas, spanning the years 2003 to 2016, with a specific focus on non-Hispanic whites and Mexican-Americans. For additional details, refer to \citet[p.~1027]{Honore_Hu:2020}.
The estimated densities are illustrated in Figures \ref{fig: wage1}--\ref{fig: wage4}, wherein the sample mean and median are also reported.
The Epanechnikov kernel and \citeauthor{Sheather_Jones:1991}'s (\citeyear{Sheather_Jones:1991}) bandwidth selector are used.
The estimated densities in Figures \ref{fig: wage1}--\ref{fig: wage3} seems symmetric. The density in Figure \ref{fig: wage4} is asymmetric while the sample mean and median are similar.

\section{Marginal Treatment Effects in the Presence of Sample Selection}
The stochastic monotonicity assumption will be applicable to many settings that are analysed under the monotonicity assumption.
In this section, as a simple but important application, we consider the identification of the marginal treatment effect (MTE) in the presence of sample selection, which is analysed by \cite{Bartalotti_etal:2023}.
In their work, the monotone sample selection as in Assumption \ref{assumptionOA: monotonicity} below plays an important role in obtaining useful bounds; they wrote, ``the identified set can be substantially tightened by imposing that the sample selection mechanism is monotone in the treatment."

Again, in the case of MTE, the monotonicity may not always hold, as discussed in the main text. 
Therefore, empirical researchers can only obtain bounds that are too large to be informative when they are concerned about the monotonicity assumption and abandon the bounds based on this assumption.
Stochastic monotonicity will be an effective alternative in such a case.

\subsection{Model, Assumptions, and Identification}
Following \cite{Bartalotti_etal:2023}, we consider the next model:
\begin{subequations}\label{mte model}
\begin{align}
    & D_i = \mathbf{1}\left\{V_i \leq P(Z_i)\right\},\\
    & S_i=S_{1i}D_i + S_{0i}(1-D_i),\\
    &Y_i = S_i\cdot\left\{Y_{1i}^* D_i + Y_{0i}^* (1-D_i)\right\},\\
    &(Y_i, S_i, D_i, Z_i)\text{ is observed},
\end{align}
\end{subequations}
where all the characters are defined in the same manner as before, except for $Z_i$ and $V_i$. 
$Z_i$ is a vector of continuous instrumental variables, and $V_i$ is a latent variable.
As in \cite{Bartalotti_etal:2023}, our target is the following MTE function for the always-takers:
\begin{align*}
    \mu(v) = \E{Y_{1i}^* - Y_{0i}^* | V = v, S_{0i}=1, S_{1i}=1}.
\end{align*}
\cite{Bartalotti_etal:2023} shows the identification result under the next set of assumptions and Assumption \ref{assumptionOA: monotonicity}, i.e., the monotonicity assumption.
\begin{assumption}[\citealp{Bartalotti_etal:2023}]\label{assumptionOA mte}
    \item[(i)] The instruments $Z_i$ are independent of $(Y_{1i}^*, Y_{0i}^*, S_{1i}, S_{0i}, V_i)$.
    \item[(ii)] $P(z)$ is a nontrivial function of $z$ and the random variable $P(Z_i)$ is absolutely continuous with support given by an interval $\mathcal{P}\coloneqq[p_L,p_U]\subseteq[0,1]$.
    \item[(iii)] $0<\P{D_i=1}<1$ and $\P{S_{0i}=1,S_{1i}=1|V=v}>0$ for any $v\in\mathcal{P}$.
    \item[(iv)] $\E{\left|Y_{di}^*\right| \big| S_{0i}=1, S_{1i}=1, V=v}<\infty$ for any $v\in\mathcal{P}$ and $d\in\{0,1\}$.
    \item[(v)] The unconditional distribution of $V_i$ is uniform over $[0,1]$.
\end{assumption}
 \begin{assumption}[Monotonicity]\label{assumptionOA: monotonicity}
    $\P{S_{1i}=1 | S_{0i}=1}=1$.
\end{assumption}

Instead of imposing this exact monotonicity assumption (Assumption \ref{assumptionOA: monotonicity}), we consider the identification under a slightly modified version of the stochastic monotonicity assumption in the main text:
\begin{assumption}[Stochastic Monotonicity for MTE]\label{assumptionOA: stochastic monotonicity for mte}
    \item[(i)] $\P{S_{1i}=1|S_{0i}=1, V_i=v}\geq\vartheta_L$ with known $\vartheta_L\in(0,1]$ for any $v\in\mathcal{P}$.
    \item[(ii)] $(Y_{1i}^*, Y_{0i}^*)$ is independent of $S_{1i}$ conditional on $S_{0i}=1, V=v$.
\end{assumption}
The intuition behind this assumption is similar to before. 
Under this version of monotonicity, we can again obtain bounds as follows:
\begin{theorem}\label{thm MTE}
    Assume the model \eqref{mte model}. Let $Y_{0i}^*$ and $Y_{1i}^*$ be continuous random variables.
    Under Assumptions \ref{assumptionOA mte} and \ref{assumptionOA: stochastic monotonicity for mte}, for any $v\in\mathcal{P}$, the upper and lower (uniformly) sharp bounds, $\nabla_{\mu}(v)$ and $\rnabla_{\mu}(v)$, for $\mu(v)$ are given by
    \begin{align*}
        \nabla_{\mu}(v) =& \E{\Tilde{Y}_{1i} | D_i = 1, S_i = 1, P(Z_i) = v, \Tilde{Y}_{1i} \geq y_{1-s(v)}^1}- \E{\Tilde{Y}_{0i} | D_i = 0, S_i = 1, P(Z_i) = v},\\
        \rnabla_{\mu}(v) =& \E{\Tilde{Y}_{1i} | D_i = 1, S_i = 1, P(Z_i) = v, \Tilde{Y}_{1i} \leq y_{s(v)}^1}- \E{\Tilde{Y}_{0i} | D_i = 0, S_i = 1, P(Z_i) = v},
    \end{align*}
    where $s(v)=\max\{\vartheta_L\eta(v), \Tilde{\alpha}(v)\}$.
    The conditional distribution of $\Tilde{Y}_{di}$ (given $D_i=d$, $S_i=1$, and $P(Z_i)=v$), written by $\Tilde{F}_d$, is given by
    \begin{align*}
        \Tilde{F}_d(y) = \left(\frac{\partial \P{Y\leq Y, S_i=1, D_i=d | P(Z_i)=v}}{\partial v}\right)\bigg/ \left(\frac{\partial \P{S_i=1, D_i=d | P(Z_i)=v}}{\partial v}\right),
    \end{align*}
    and $y_r^1 = \Tilde{F}_1^{-1} (r)$.
    The functions $\eta$ and $\Tilde{\alpha}$ are given by
    \begin{align*}
        \eta(v) = -\left(\frac{\partial \P{S_i=1, D_i=0 | P(Z_i)=v}}{\partial v}\right)\bigg/ \left(\frac{\partial \P{S_i=1, D_i=1 | P(Z_i)=v}}{\partial v}\right),
    \end{align*}
    and 
    \begin{align*}
        \Tilde{\alpha}(v) = \max\left\{-\frac{\partial \P{S_i=1, D_i=0 | P(Z_i)=v}}{\partial v}+\frac{\partial \P{S_i=1, D_i=1 | P(Z_i)=v}}{\partial v}-1,0\right\}\\
        \bigg/ 
        \left(\frac{\partial \P{S_i=1, D_i=1 | P(Z_i)=v}}{\partial v}\right).
    \end{align*}
\end{theorem}

\noindent
\textbf{Proof of Theorem \ref{assumptionOA: stochastic monotonicity for mte}:}
We first summarize for convenience important identification results that hold under Assumption \ref{assumptionOA mte} derived in \citet[Section 3.1]{Bartalotti_etal:2023}.
Let $\mathcal{Y}\subseteq\mathbb{R}$ be the support of $Y_{0i}^*$ and $Y_{1i}^*$.
For any $v\in\mathcal{P}$ and any Borel set $\mathcal{A}\subseteq\mathcal{Y}$, 
\begin{align}
    \P{Y_{1i}^*\in\mathcal{A}, S_{1i}=1|V_i=v} &= \frac{\partial \P{Y_i\in\mathcal{A},S_i=1,D_i=1|P(Z_i)=v}}{\partial v},\label{OA eq1 MTE proof}\\
    \P{Y_{0i}^*\in\mathcal{A}, S_{0i}=1|V_i=v} &= -\frac{\partial \P{Y_i\in\mathcal{A},S_i=1,D_i=0|P(Z_i)=v}}{\partial v}.\label{OA eq2 MTE proof}
\end{align}
Also, we immediately have that
\begin{align}
    \P{S_{1i}=1|V_i=v} &= \frac{\partial \P{S_i=1,D_i=1|P(Z_i)=v}}{\partial v},\label{OA eq3 MTE proof}\\
    \P{S_{0i}=1|V_i=v} &= -\frac{\partial \P{S_i=1,D_i=0|P(Z_i)=v}}{\partial v}.\label{OA eq4 MTE proof}
\end{align}

Given the above relations, we can show that under Assumptions \ref{assumptionOA mte} and \ref{assumptionOA: stochastic monotonicity for mte}, it holds that
\begin{align*}
    \E{Y_{0i}^* | V_i=v, S_{0i}=1, S_{1i}=1} 
    = \E{\Tilde{Y}_{0i} | S_i=1, D_i=0, P(Z_i)=v}.
\end{align*}
Hence, we will consider the bounds for $\E{Y_{1i}^* | V_i=v, S_{0i}=1, S_{1i}=1}$.

Firstly, by \citet[equation (3.7)]{Bartalotti_etal:2023}, it holds that
\begin{align*}
    \P{Y_{1i}^* \in\mathcal{A}| S_{1i}=1, V_i=v} =& \alpha(v) \P{Y_{1i}^* \in\mathcal{A}| S_{0i}=1, S_{1i}=1, V_i=v}\\
    &+ \left(1-\alpha(v)\right) \P{Y_{1i}^* \in\mathcal{A}| S_{0i}=0, S_{1i}=1, V_i=v},
\end{align*}
for any Borel set $\mathcal{A}\subseteq\mathcal{Y}$, where
\begin{align*}
    \alpha(v) = \frac{\P{S_{1i}=1, S_{0i}=1 | V=v}}{\P{S_{1i}=1|V=v}}.
\end{align*}

By Lemma 1 of \cite{Bartalotti_etal:2023}, the Fréchet inequalities produce a sharp lower bound $\alpha(v) \geq \Tilde{\alpha}(v)$ under Assumption \ref{assumptionOA mte}.
Also, Assumption \ref{assumptionOA: stochastic monotonicity for mte} and \eqref{OA eq3 MTE proof}-\eqref{OA eq4 MTE proof} imply that $\alpha(v) \geq \vartheta_L\eta(v)$.
Therefore, for any $v$, $\alpha(v)\geq s(v)\coloneqq\max\{\vartheta_L\eta(v), \Tilde{\alpha}(v)\}$. 
Thus, Proposition 4 of \cite{Horowitz_Manski:1995} implies that the bounds $[\rnabla_\mu(v),\nabla_\mu(v)]$ are valid.

The uniform sharpness can be shown in a similar way to Proposition 1 of \cite{Bartalotti_etal:2023, Bartalotti_etal:2023_appendix}.
Define $s_N(v) = \max\{\vartheta_L\eta_N(v), \Tilde{\alpha}_N(v)\}$ where
\begin{align*}
    \eta_N(v) &= -\left(\frac{\partial \P{S_i=1, D_i=0 | P(Z_i)=v}}{\partial v}\right),\\
    \Tilde{\alpha}_N(v) &= \max\left\{-\frac{\partial \P{S_i=1, D_i=0 | P(Z_i)=v}}{\partial v}+\frac{\partial \P{S_i=1, D_i=1 | P(Z_i)=v}}{\partial v}-1,0\right\},
\end{align*}
i.e., the ``$N$"umerator of $\eta(v), \Tilde{\alpha}(v)$.
Below, we show that the lower bound $\rnabla_\mu$ is achievable. The upper bound is similar.
Following \citet[p.~7]{Bartalotti_etal:2023_appendix}, define $\Tilde{V}\sim\mathrm{Uniform}[0,1]$, and 
\begin{align*}
    k(v) \coloneqq \begin{cases}
        s_N(v) & \text{ if } v\in\mathcal{P}\\
        \epsilon_0  & \text{ if } v\in[0,p_L)\\
        \epsilon_1  & \text{ if } v\in(p_U,1]
    \end{cases},
\end{align*}
where $\epsilon_0,\epsilon_1\in[0,1]$ are some appropriate constants.
Define 
\begin{align*}
    \P{\Tilde{S}_{1i}=s_1 | \Tilde{V}_i=v} &\coloneqq \begin{cases}
        \dfrac{\partial \P{S_i = s_1, D_i = 1 | P(Z_i) = v}}{\partial v}& \text{ if } v\in\mathcal{P}\\
        \dfrac{\P{S_i = s_1, D_i = 1 | P(Z_i) = p_L}}{p_L} & \text{ if } v\in[0,p_L)\\
        k(v)\mathbf{1}\{s_1=1\} + (1-k(v))\mathbf{1}\{s_1=0\} & \text{ if } v\in(p_U,1]
    \end{cases}\\
    \P{\Tilde{S}_{0i}=s_0 | \Tilde{V}_i=v} &\coloneqq \begin{cases}
        -\dfrac{\partial \P{S_i = s_0, D_i = 0 | P(Z_i) = v}}{\partial v}& \text{ if } v\in\mathcal{P}\\
        k(v)\mathbf{1}\{s_0=1\} + (1-k(v))\mathbf{1}\{s_0=0\} & \text{ if } v\in[0,p_L)\\
        \dfrac{\P{S_i = s_0, D_i = 0 | P(Z_i) = p_U}}{1-p_U}& \text{ if } v\in(p_U,1]
    \end{cases},
\end{align*}
where $s_1,s_0\in\{0,1\}$.
Then, defining $\pi_{11}\coloneqq\P{\Tilde{S}_{0i} = 1 , \Tilde{S}_{1i} = 1 | \Tilde{V}_i=v} \coloneqq k(v)$, we have
\begin{align*}
    \pi_{01}&\coloneqq\P{\Tilde{S}_{0i} = 0 , \Tilde{S}_{1i} = 1 | \Tilde{V}_i=v} =
    \P{\Tilde{S}_{1i}=1|\Tilde{V}_i=v} - k(v),\\
    \pi_{10}&\coloneqq\P{\Tilde{S}_{0i} = 1 , \Tilde{S}_{1i} = 0 | \Tilde{V}_i=v} =
    \P{\Tilde{S}_{0i}=1|\Tilde{V}_i=v} - k(v),\\
    \pi_{00}&\coloneqq\P{\Tilde{S}_{0i} = 0 , \Tilde{S}_{1i} = 0 | \Tilde{V}_i=v} =
    1 - \left(\pi_{11} + \pi_{01} + \pi_{10} \right),
\end{align*}
these are in $[0,1]$ by Assumption \ref{assumptionOA: stochastic monotonicity for mte} and construction. We can see that Assumption \ref{assumptionOA: stochastic monotonicity for mte}-(i) is satisfied.
Define
\begin{align*}
    &\P{\Tilde{Y}_{1i}^*\leq y_1, \Tilde{S}_{1i}=s_1 | \Tilde{V}_i=v}\\ &\quad\coloneqq \begin{cases}
        \dfrac{\partial \P{Y_i\leq y_1, S_i = s_1, D_i = 1 | P(Z_i) = v}}{\partial v}& \text{ if } v\in\mathcal{P}\\
        \dfrac{\P{Y_i\leq y_1, S_i = s_1, D_i = 1 | P(Z_i) = p_L}}{p_L} & \text{ if } v\in[0,p_L)\\
        \left\{k(v)\mathbf{1}\{s_1=1\} + (1-k(v))\mathbf{1}\{s_1=0\}\right\}\P{Y_i\leq y_1} & \text{ if } v\in(p_U,1]
    \end{cases}\\
    &\P{\Tilde{Y}_{0i}^*\leq y_0, \Tilde{S}_{0i}=s_0 | \Tilde{V}_i=v}\\ &\quad\coloneqq
    \begin{cases}
        -\dfrac{\partial \P{Y_i\leq y_0, S_i = s_0, D_i = 0 | P(Z_i) = v}}{\partial v}& \text{ if } v\in\mathcal{P}\\
        \left\{k(v)\mathbf{1}\{s_0=1\} + (1-k(v))\mathbf{1}\{s_0=0\}\right\}\P{Y_i\leq y_0} & \text{ if } v\in[0,p_L)\\
        \dfrac{\P{Y_i\leq y_0, S_i = s_0, D_i = 0 | P(Z_i) = p_U}}{1-p_U}& \text{ if } v\in(p_U,1]
    \end{cases},
\end{align*}
which are proper distribution functions by \eqref{OA eq1 MTE proof}, \eqref{OA eq2 MTE proof}.
Hence, letting $\Tilde{Y}_{0i} \sim F_{\Tilde{Y}_0^*|\Tilde{S}_0=1,\Tilde{V}=v}$ and $\Tilde{Y}_{1i} \sim F_{\Tilde{Y}_1^*|\Tilde{S}_1=1,\Tilde{V}=v}$, we can define
\begin{align*}
    \P{\Tilde{Y}_{1i}^* \leq y_1 | \Tilde{S}_{0i}=1, \Tilde{S}_{1i}=s_1, \Tilde{V}=v} &\coloneqq 
    \P{\Tilde{Y}_{1i} \leq y_1 | \Tilde{Y}_{1i}\leq F^{-1}_{\Tilde{Y}_1}\left(\frac{\pi_{11}}{\pi_{11} + \pi_{01}}\right) },\\
    \P{\Tilde{Y}_{1i}^* \leq y_1 | \Tilde{S}_{0i}=0, \Tilde{S}_{1i}=s_1, \Tilde{V}=v} &\coloneqq
    \P{\Tilde{Y}_{1i} \leq y_1 | \Tilde{Y}_{1i} > F^{-1}_{\Tilde{Y}_1}\left(\frac{\pi_{11}}{\pi_{11} + \pi_{01}}\right) },\\
    \P{\Tilde{Y}_{0i}^* \leq y_0 | \Tilde{S}_{0i}=s_0, \Tilde{S}_{1i}=s_1, \Tilde{V}=v} &\coloneqq \P{\Tilde{Y}_{0i}\leq y_0},
\end{align*}
for $\forall s_1,s_0\in\{0,1\}$. 
Then, Assumption \ref{assumptionOA: stochastic monotonicity for mte}-(ii) is satisfied.
Define $\Tilde{D}_i = \mathbf{1}\{\Tilde{V}_I \leq P(Z_i)\}$, then we can see that
\begin{align*}
    \P{\Tilde{Y}_i \leq y, \Tilde{S}_i = 1, \Tilde{D}_i = 1 | Z_i = z} 
    &=
    \P{\Tilde{Y}_{1i}^* \leq y, \Tilde{S}_{1i} = 1, \Tilde{V}_i \leq P(z)} \\
    &=
    \int_0^{P(z)} \P{\Tilde{Y}_{1i}^* \leq y, \Tilde{S}_{1i} = 1 | \Tilde{V}_i = v}\, dv \\
    &=
    \P{Y_i \leq y, S_i = 1, D_i = 1 | P(Z_i) = P(z)}\\
    &= 
    \P{Y_i \leq y, S_i = 1, D_i = 1 | Z_i=z},
\end{align*}
where the third equality uses the above definition of the integrand; for the second equality, see \citet[p.~570]{Bartalotti_etal:2023}.
A similar argument holds for $\mathbb{P}[\Tilde{Y}_i \leq y, \Tilde{S}_i = 1, \Tilde{D}_i = 0 | Z_i = z]$.
Hence, the defined set of variables with tildes induces the joint distribution of the observed data.
Finally, we can obtain that, for $v\in\mathcal{P}$,
\begin{align*}
    \rnabla_\mu(v) = \E{\Tilde{Y}_{1i}^* - \Tilde{Y}_{0i}^* | \Tilde{S}_{0i}=1, \Tilde{S}_{1i}=1, \Tilde{V}=v},
\end{align*}
that is, the lower bound is attainable.\hfill$\square$\\

\begin{remark}
    The uniform sharpness means that for any function $f:\mathcal{P}\to\mathbb{R}$ such that $f\in[\rnabla_\mu, \nabla_\mu]$, we can construct random variables that are consistent with the model and assumptions, imply the joint distribution of observables is the same as the observed distribution, and produce $f$ as the MTE function. See \citet[Remark 3]{Bartalotti_etal:2023}. 
\end{remark}
Obviously, (i) Assumption \ref{assumptionOA mte}$+$Assumption \ref{assumptionOA: monotonicity} $\Longrightarrow$ (ii) Assumption \ref{assumptionOA mte}$+$Assumption \ref{assumptionOA: stochastic monotonicity for mte} $\Longrightarrow$ (iii) Assumption \ref{assumptionOA mte}.
\cite{Bartalotti_etal:2023} derive identification under (i) and (iii), and find that the bounds under (iii) are too wide to be informative \citep[p.~29]{Bartalotti_etal:2023_appendix}.
Hence, the bounds under (ii), which are obtained in Theorem \ref{thm MTE}, will be much more informative than the bounds under (iii) but are more robust than (i).

\begin{example}[Numerical Example]
We here compute the MTE bounds using the same data generating process (DGP) as in \citet[p.~29]{Bartalotti_etal:2023_appendix}.
Consider the following DGP, which is adopted from \citet[p.~29]{Bartalotti_etal:2023_appendix}:
\begin{align*}
    &D = \mathbf{1}\left\{V \leq \Phi(Z)\right\},\, 
    S = \mathbf{1}\left\{U_S \leq 0.1 + 0.4 D\right\},\\ 
    &Y^* = Y_{1}^*D + Y_{0}^*(1-D), \, 
    Y = Y^* S,
\end{align*}
where $V=\Phi(\epsilon_V)$, $U_S=(\epsilon_V + \epsilon_S)/\sqrt{2}$, $(\epsilon_V,\epsilon_S, Z, \xi)\sim \mathcal{N}(0,I)$, $I$ is the identity matrix, $\Phi$ is the distribution function of $\mathcal{N}(0,1)$, and
\begin{align*}
    &Y_0^* = \mathbf{1}\{\xi\geq0\}\cdot \epsilon_V + \mathbf{1}\{\xi<0\}\cdot (-\epsilon_V),\\
    &Y_1^* = \mathbf{1}\{\xi\geq0\}\cdot 5\epsilon_V + \mathbf{1}\{\xi<0\}\cdot (-\epsilon_V).
\end{align*}
The bounds under (i) Assumption \ref{assumptionOA mte}$+$Assumption \ref{assumptionOA: monotonicity}, (ii) Assumption \ref{assumptionOA mte}$+$Assumption \ref{assumptionOA: stochastic monotonicity for mte}, and (iii) Assumption \ref{assumptionOA mte} are shown in Figure \ref{OA fig: mte}. Note that we set $\vartheta_L=0.8$ to obtain visually clear bounds, and we use numerical integration to compute bounds.
Note that all sets of assumptions are valid in this example.
The solid green line is the bound under the stochastic monotonicity (i.e., (ii) with $\vartheta_L=0.8$).
The bounds are, of course, wider than the bounds under (i), while the bounds are much more informative than those under (iii).
Therefore, Assumption \ref{assumptionOA: stochastic monotonicity for mte} will be useful when researchers lack confidence in the validity of the monotonicity assumption.
\begin{figure}[th]
    \centering
    \includegraphics[scale=0.7]{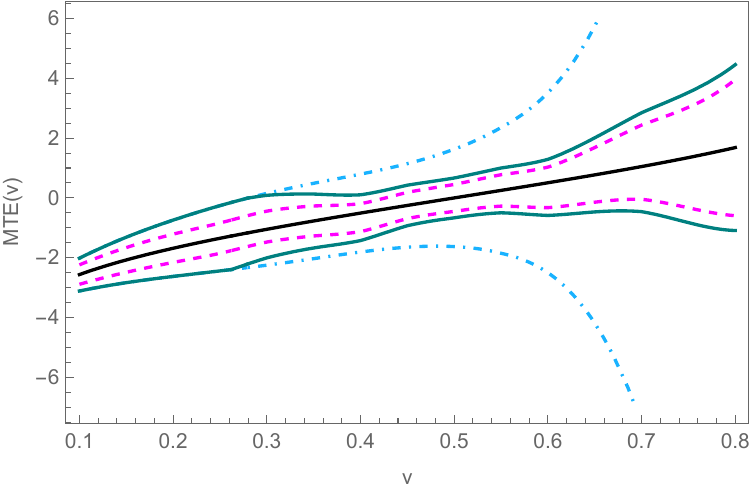}
    \caption{Bounds for $\mu(v)$ under Several Sets of Assumptions. \textit{The dashed red lines are bounds under (i), the solid green lines are bounds under (ii) with $\vartheta_L=0.8$, and the dot-dashed blue lines are bounds under (iii). The black solid line in the centre is the true $\mu(v)$.}}
    \label{OA fig: mte}
\end{figure}
\end{example}

\end{document}